\documentclass[12pt]{report}\usepackage[letterpaper,top=1in, bottom=1in, left=1.5in, right=1in]{geometry}
\usepackage{amsmath,amssymb,leftidx,mathtools}
\usepackage{graphicx,tikz}
\usepackage{caption, subcaption, float}
\usepackage{authblk}
\usepackage{hyperref}
\usepackage[utf8]{inputenc}
\usepackage[english]{babel}
\addto{\captionsenglish}{}
\usepackage[strict]{changepage}
\usepackage{setspace}

\usepackage{chngcntr}
\counterwithout{footnote}{chapter}

\usepackage{fancyhdr}

\usepackage{xcolor}
\hypersetup{
    colorlinks,
    linkcolor={black!50!black},
    citecolor={black!50!black},
    urlcolor={black!80!black}
}

\usepackage{csquotes}

\usepackage{stackengine}

\usepackage{calligra}

\DeclareMathAlphabet{\mathcalligra}{T1}{calligra}{m}{n}
\DeclareFontShape{T1}{calligra}{m}{n}{<->s*[2.2]callig15}{}
\newcommand{\scripty}[1]{\ensuremath{\mathcalligra{#1}}}

\renewcommand\maketitle{}

\begin{document}

%\title{Green Functions, Sommerfeld Images, and Wormholes}

%\author{Hassan Alshal}

%\affil{Department of Physics, University of Miami}
\begin{titlepage}
   \begin{center}
       \vspace*{2cm}

		UNIVERSITY OF MIAMI 

       \vspace*{2cm}
 
       GREEN FUNCTIONS, SOMMERFELD IMAGES, AND WORMHOLES
 
       \vspace{2cm}
		By 
       \vspace{2cm}
 
       HASSAN ALSHAL
       
        \vspace{2cm}

        A THESIS

        \vspace{4cm}

Submitted to the Faculty\\
of the University of Miami\\
in partial fulfillment of the requirements for\\
the degree of Master of Science\\

       \vfill

Coral Gables, Florida\\
May 2019
 
   \end{center}
\end{titlepage}

\pagestyle{myheadings}
\lhead{}

\maketitle
\thispagestyle{empty}

\renewcommand{\abstractname}{\flushleft\LARGE{Abstract}\hfill}
\begin{abstract}
\doublespacing

\vspace*{0.5cm}

\noindent
Thesis supervised by Professor Thomas L. Curtright.\\
\vspace*{-0.5cm}

\noindent Electrostatic Green functions for grounded equipotential circular and elliptical rings, and grounded hyperspheres in n-dimension electrostatics, are constructed using Sommerfeld’s method. These electrostatic systems are treated geometrically as different radial p-norm wormhole metrics that are deformed to be the Manhattan norm, namely “squashed wormholes”. Differential geometry techniques are discussed to show how Riemannian geometry plays a rule in Sommerfeld’s method. A comparison is made in terms of strength and position of the image charges for Sommerfeld’s method with those for the more conventional Kelvin’s method. Both methods are shown to be mathematically equivalent in terms of the corresponding Green functions. However, the two methods provide different physics perspectives, especially when studying different limits of those electrostatic systems. Further studies of ellipsoidal cases are suggested.
\end{abstract}

\renewcommand{\abstractname}{}
\begin{abstract}
\vspace{3cm}

In tribute to Richard Feynman (1918-1988) and Arnold Sommerfeld (1868-1951).
\end{abstract}

\renewcommand{\abstractname}{\flushleft\LARGE{Acknowledgements}\hfill}
\begin{abstract}
\doublespacing
\vspace*{0.5cm}

I would like to show my deep gratitude to Prof. Dr. Thomas L. Curtright who has never shown parsimony or hesitance to support and advise me through education and research, either in this thesis or in any other curricular endeavors. I believe his leading personality together with his wide knowledge of the topic were indispensable factors for me to finish this thesis. I am forever indebted to him.
\flushright HASSAN ALSHAL
\flushleft University of Miami\\
May 2019
\end{abstract}

\doublespacing
\tableofcontents

\setcounter{page}{2}
\addcontentsline{toc}{chapter}{List of Figures}
%\begingroup
%\setstretch{1.5}

\chapter*{List of Figures}

Figure 1.a: Frontal view of extended real coordinates geometry to Kelvin's method for grounded ellipse (red) with $U=3/2$. The \textquote{doorway} (blue) is in the middle between  exterior charges (orange) and corresponding images (green) regions. Notice the phase change in the image trajectories $\Delta v=2 v'$ when they cross the ``doorway'' at $u=0$. Grey curves are parameterized ellipses with fixed $u$ and $0\leq v \leq 2\pi$...............................................................................................................38
%\tikz \draw[loosely dotted] (0pt,0pt) -- (285pt,0pt); 33
\\
\\
Figure 1.b: Lateral view of of extended real coordinates geometry to Kelvin's method with for side view of grounded ellipse (red line) with $U=3/2$ where charges and images trajectories (orange) and (green) are shown respectively. Notice the phase change in the image trajectories $\Delta v=2 v'$ when they cross the ``doorway'' at $u=0$. Grey curves are parameterized ellipses with fixed $u$ and $0\leq v \leq 2\pi$...................................39
%\tikz \draw[loosely dotted] (0pt,0pt) -- (285pt,0pt); 34
\\
\\
Figure 2.a: Bird view of real coordinates geometry of Sommerfeld's method for grounded ellipse (red) at $U=0$ with trajectories of exterior sources (orange) and their corresponding images (green). In contrary to Kelvin's method, the ``doorway'' is the ring itself. Grey curves are parameterized ellipses with fixed $u$ and $0\leq v \leq 2\pi$...............................................................................................................40
%\tikz \draw[loosely dotted] (0pt,0pt) -- (285pt,0pt); 35
\\
\\
\\
Figure 2.b: Frontal view of real coordinates geometry of Sommerfeld's method for grounded ellipse (red) at $U=0$. Unlike Kelvin's method, Sommerfeld's method fixes trajectories of exterior sources (orange) and their corresponding images (green) at same angle $v$. Grey curves are parameterized ellipses with fixed $u$ and $0\leq v \leq 2\pi$ .................................................................................................................................41
%\tikz \draw[loosely dotted] (0pt,0pt) -- (285pt,0pt); 36
%\endgroup

\addcontentsline{toc}{chapter}{Preface}
\chapter*{Preface}

\hspace{\parindent} When George Green introduced his technique of Green functions, he didn't realize he had changed the face of both mathematics and physics forever. Being inspired by work of Pierre-Simon Laplace and Sim\'eon Poisson on electric potentials\footnote {Indeed Green coined the term \textit{Potential Theory} \cite{Ggm}.} \cite{Ggf}, Green introduced the concept of Green functions in his seminal work: \textquotedblleft \textit{An Essay on the Application of mathematical Analysis to the theories of Electricity and Magnetism.}\textquotedblright As he was aware he lacked an institutional scientific degree, Green was hesitant to submit it for publication in a journal of those managed by the only two scientific societies in England at that time: British Royal Society and Cambridge Philosophical Society. Instead he printed it at his own expense and distributed it among his fellows in Nottingham Subscription Library society. As it stayed unpublished in any journal even after Green earned the Tripos degree from Cambridge and passed away two years later, his essay didn't capture the attention of professional mathematicians until William Thomson \cite{Tnp}, later Lord Kelvin, shared it with French mathematicians like Chasles, Liouville and Sturm, during his stay in Paris, and the French mathematicians were astounded by Green's work. According to Kelvin, Sturm expressed his admiration: \textquote{Ah voil\`a mon affaire!} [Oh that's my business!] as he found what he independently discovered was originally established by Green 16-17 years before him. After he returned home, Kelvin managed to republish it in Crelle's journal (Berlin) in three separate parts in 1850, 1852, and 1854 \footnote{It has been republished recently on \href{https://arxiv.org/abs/0807.0088}{[arXiv:0807.0088].}}.\\

In fact the \textquote{revolutionary} \footnote{was introduced in Spring of Nations year; Thomson W., Geometrical Investigations with Reference to the Distribution of Electricity on Spherical Conductors, Camb. Dublin Math. J. 3, 141 (1848).} Kelvin's method of images is a direct application of Green functions in electrostatics problems. Also Green's work inspired George Stokes\cite{Ggm}, a close Cambridgean friend of Kelvin, while he was persuing his study to wave phenomena in hydrodynamical systems.\\
Away for being known for introducing another set of conditions related to Green functions, Carl Neumann \cite{{Heh}} focused on potential theory in 2D using Dirichlet boundary conditions. He eventually came up with a solution he called \textit{Logarithmischen Potential}, which will be discussed in chpater 2 when the potential problem with a grounded 2D conducting ring is treated. Other German mathematicians \cite{Gfa} extensively studied Green function techniques, meanwhile Bernhard Riemann \cite{Ggm} was the one who coined its name.\\
By 1880s-90s, Green functions became a hot topic specially among theoretical and mathematical physicists who were interested in developing electromagnetic theory that could fully adopt the concept of \textquote{Luminiferous Ether} \cite{Gfa, Ase, Hta}. Being also interested in studying diffraction patterns of electromagnetic and acoustic waves, Arnold Sommerfeld generalized Kelvin's method of images to study Green functions in 2D using the concept of complex Riemann surfaces \cite{Asm, Uvp}, which is the main theme of this thesis. General consideration on how Sommerfeld built his method will be presented in the introduction chapter of this thesis.\\
Sommerfeld realized how powerful his technique is, so he wrote to Kelvin \cite{Ase} that:
\begin{displayquote}
\textquote{The number of boundary value problems solvable by means of my elaborated Thomson's method of images is very great.}
\end{displayquote}
As Whittaker \cite{Hta} described Green: \begin{displayquote}
\textquote{the founder of that \textit{Cambridge School} of natural philosophers of which Kelvin, Stokes, Rayleigh, Clerk Maxwell, Lamb, Larmor and Love were the most illustrious members in the later half of the nineteenth century},
\end{displayquote}
perhaps it would not be inappropriate to describe him as \textquotedblleft Desert father\textquotedblright \ of the \textit{propagators} business in quantum electrodynamics that was introduced by Richard Feynman, Julian Schwinger and Sin-Itiro Tomonaga \cite{Ggf} and were awarded Nobel prize on it 1965. We can see that clearly in Schwinger's lecture \cite{Sqd} as he reminisced about how he, and Feynman independently, developed the methodology of Green to construct their relativistic quantum theory of electrodynamics.\\

This thesis is devoted to study Sommerfeld's method to find Green functions of $n$D spherical and 2D elliptic \textquote{static} \textit{Ellis wormholes} \cite{Eft}, then to squash them into the so-called Manhattan norm. It is based on the following papers:
\begin{enumerate}
\item[i.] T Curtright, H Alshal, P Baral, S Huang, J Liu, K Tamang, X Zhang, and Y Zhang. \textquote{The Conducting Ring Viewed as a Wormhole} \href{https://arxiv.org/abs/1805.11147}{arXiv: 1805.11147 [physics.class-ph]}.
\item[ii.] H Alshal and T Curtright, \textquote{Grounded Hyperspheres as Squashed Wormholes} \href{https://arxiv.org/abs/1806.03762}{arXiv: 1806.03762 [physics.class-ph]}.
\item[iii.] H Alshal, T Curtright, S Subedi, \textquote{Image Charges Re-Imagined} \href{https://arxiv.org/abs/1808.08300}{arXiv: 1808.08300 [physics.class-ph]}.
\end{enumerate}
By incorporating Riemannian geometry with Sommerfeld's method, I hope this thesis convinces the reader that Sommerfeld's method, despite being not widely admired, could be considered as a precursor to studies of static wormholes \cite{Wsi, Viw}.

\chapter{Introduction}
\pagenumbering{arabic}

\hspace{\parindent}	Method of Green functions is a robust technique to obtain solutions of many linear, ordinary or partial, differential equations describing physical phenomena related to either classical or quantum field theory \cite{Ggf,Gfa}. Green functions can be seen as a response of a physical system to an impulse, e.g; a point like source of gravitational or electric field. For such physical systems the equation of motion is usually on the form $L \ [ y(x) ]=f(x)$, where $L$ is the linear operator that the corresponding Green function expresses the integral kernel of $L^{-1}$ , i.e., $G(x,x') = \mathcal{F}_{k}^{-1} \lbrace 1 / \mathcal{F}_{x} [L] \rbrace $ . And $\displaystyle{f(x)= \mathcal{F}_{k}^{-1} \left[\frac{\mathcal{F}_x [y(x)] \ (k)}{\mathcal{F}_x [G(x,x')](k)}\right]}$ which can be expressed as a Fourier series, where $\mathcal{F}_x$ and $\mathcal{F}_k^{-1}$ are the direct and inverse Fourier transforms. So from integral equations point of view, a Green function is an integral kernel of the inverse of a given operator such that Green function satisfies $L \ [ G(x,x') ]=\delta(x-x')$ where $\delta(x-x')$ is the Dirac-delta function.\\
	
	For inhomogeneous partial differential equations, i.e., with $\lambda y(x)$ not being a solution for $L[ \ y(x) \ ]=f(x)$ for some $\lambda \in \mathbb{C}$, equations mainly come with three types of boundary conditions (b.c) related to the surface $S$ that bounds the volume $V$ containing the impulse sources. Inhomogeneity comes either from b.c's, i.e., for $ y(x)=\alpha$ or $y'(x)=\beta$ where $\alpha, \beta$  are either $\neq 0$ or vary from point to another in $S$, or from the nature of the equation of motion itself. And those conditions are:
\begin{enumerate}
	\item[i.]Dirichlet boundary conditions, where $y(x)=\alpha$ for some $\alpha \in \mathbb{C}$ and $ x \in S = \partial V$. $\alpha$ could be a function in $x$.
	\item[ii.]Neumann boundary conditions, where the flux, $\displaystyle{\partial_{\textbf{\^{n}}} y(x) = \frac{\partial y(x)}{\partial \textbf{\^{n}}} = \nabla y(x) . \ \textbf{\^{n}}=\alpha}$ for some $\alpha \in \mathbb{C}$, is in the direction of the normal unit vector $\textbf{\^{n}}$ at $ x \in S = \partial V$.
	\item[iii.]Robin boundary conditions, which is a combinition of both Dirichlet and Neumann b.c's on the same boundary $S=\partial V$ \footnote{It's different from the mixed b.c's in that the mixed ones have Dirichlet and Neumann b.c's on mutually exclusive boundaries, i.e., $S_1 \cap S_2 = \phi $ and $S_1 \cup S_2 = \partial V$.}.
\end{enumerate}

\noindent For $n$-order linear operator $L$ we integrate $L \ [ G(x,x') ]=\delta(x-x')$ over period $x' \in [x' - \epsilon, x' + \epsilon]$ with $\epsilon \to 0$, we get $\displaystyle{\frac{d^{n-2} G}{d x^{n-2}} \Bigr|_{x=x'} = 0}$, while $\displaystyle{\frac{d^{n-1} G}{d x^{n-1}} \Bigr|_{x=x'} = c_n}$, i.e., the $(n-1)^{\underline{th}}$-order of differentiation of a Green function has discontinuity $ \forall x \in S $, and $c_n$ is the inverse of the coefficient of the $n^{\underline{th}}$ term of the differential equation.\\
	
	Part of potential theory asserts that for an electric charge within the vicinity of an idealized conductor, a geometric distribution of \textit{induced} charge is produced to be restricted on the surface of the that conductor. In ideal electrostatic circumstances the \textit{interior} of the conductor can not contain any induced charge as it's a physical equipotential volume. However, for the sake of finding the effect of the induced charge on the surface, it would be expedient to \textit{imagine} as if the surface is neutrally like a mirror reflecting \textit{images} of another charge distribution of opposite sign to the original charge \footnote{In general nothing precludes both charge and its image(s) to be on the same side of the conductor surface depending on the geometry of the problem.}. That \textit{image} charges are \emph{not} uniquely dictated although they are endowed with electrostatic effects similar to those of the actual distribution of induced surface charge. The mutual effect of both charge and its image is determined only based on the surface of the conductor, namely, the \textit{boundary}. There are two geometrical techniques governing the mirror image methods: the well-known conventional Kelvin's method and Sommerfeld's method. Those two methods render different geometrical image regions despite they yield exactly the same results. This geometrical freedom is only restricted to the boundary determined by the surface separating the exterior and the interior of the conductor. In general, Kelvin's method is preferable when it is manageable to extend Green functions from the charge domain to the image domain geometries, while Sommerfeld's method is recommended for conductor surfaces producing images at obvious conformational positions similar to the pattern of reflections of the \textit{real} world on a flat mirror.\\
	
Sommerfeld's method imposes a split duplicate 2D space laying over the original one \cite{Asm}. Both spaces are endowed with Riemann surfaces properties. A Riemann surface is 2D real manifold (surface) with complex structure to adopt complex-analytic functions. The two spaces are separated through \textit{branch cuts} which intersect the grounded surface, where the impulse source of waves is forbidden to exist on, at its boundaries. For the upper double space a point $x_{up}$ with $\theta \in [0,2\pi)$ while for the lower one a point $x_{down}$ has $\theta \in [2\pi,4\pi)$ such that each point on a space has correspondent coincident point upon squashing \textit{double-sheeted Riemann surface}. Then the mirror image of the source on the lower space would be away from the boundary at distance exactly equal to that of the source on the upper space.\\
	Due to conformational symmetry  corresponding to such boundary-value problem, Sommerfeld expected the behavior of the potential function of the source and the image separately to be identical up to a sign and angles difference in both double spaces. Meanwhile the difference in both functions that represent the actual potential, is \emph{not} equal to zero, except for those points of the grounded surface with $\theta \in \lbrace 0, 2\pi \rbrace$ \footnote{Check figures [6-13] p.69-77 in \cite{Asm}.} as branching locates there.\\
	Later Sommerfeld \cite{Uvp} used his generalization to Kelvin's method to study the potential of point charge near a grounded 2D disc by considering Green's second identity which states that for harmonic and continuously differentiable functions $u(x)$ and $v(x)$ in volume $V$ bounded by surface $\partial V = S$ and the normal unit vector of that surface is $\textbf{\^{n}}$ we have:
	\begin{equation} \label{eq.1.1}
	\displaystyle{ \int \limits_{V} \left[ u(x) \nabla^2 v(x) - v(x) \nabla^2 u(x) \right] \ dV = \int \limits_{S=\partial V}  \left[ u(x) \nabla v(x) - v(x) \nabla u(x) \right] \cdot \textbf{\^{n}} \ dS}~.
	\end{equation}
So for any $2^{\underline{nd}}$ order linear operator $L$, a Green function obeys b.c's of $y(x)$ together with:
\begin{equation} \label{eq.1.2}
	\displaystyle{G(x,x')\Bigr|_{x=x'}}=0, \ \textnormal{and a discontinuity:} \  \displaystyle{\frac{\partial G(x,x')}{\partial {x'}} \Bigr|_{x=x'}=c}.
\end{equation}
Then by applying Green's second identity to Poisson's equation $\nabla^2 u(\vec{r})=\varrho(\vec{r})/\epsilon_0$ and using suitable Dirichlet b.c's: $ \forall \vec{r} \in S, \ u(\vec{r})=f(\vec{r})$ and $G(\vec{r},\vec{r} \, ')=0$, Sommerfeld obtained:
	\begin{equation} \label{eq.1.3}
	u(\vec{r})= \int \limits_V G(\vec{r},\vec{r} \ ') \ \varrho(\vec{r} \ ') \ dr'^3 + \int \limits_S f(\vec{r} \ ') \ \frac{\partial G(\vec{r},\vec{r} \ ')}{\partial n} \ dr'^2 \, .
\end{equation}
In Sommerfeld treatment, had b.c's been homogeneous, the surface integral would have vanished. Also $G(\vec{r}, \vec{r} \ ')$ vanishes at $S$ together with $\nabla^2 [G(\vec{r}, \vec{r} \ ')]=\delta(\vec{r}-\vec{r} \ ')$. Then Green function can be separated into \textit{fundamental solution} $G(\vec{r}, \vec{r} \ ')$ and its  \textquotedblleft copies\textquotedblright \ $G_i(\vec{r}, \vec{r_i} \ ')$ that render \textit{mirror image solutions} of the fundamental solution outside $V$ in $V_i$'s for some $i$ , That is:
\begin{equation} \label{eq.1.4}
G_o(\vec{r}, \vec{r} \, ') = G(\vec{r}, \vec{r} \, ') + \sum \limits_{i=1}^N G_i(\vec{r}, \vec{r_i} \, ') \, .
\end{equation}
Both fundamental and mirror image solutions don't necessarily satisfy the b.c's. What actually matters is adjusting them such that: 
\begin{itemize}

\item[] a) \ $\nabla^2 [G(\vec{r}, \vec{r} \, ')]=\delta(\vec{r}-\vec{r} \, ')~\cdot$ 
\item[] b) \ $\nabla^2 \ G_i(\vec{r}, \vec{r_i} \, ')=0 \ \forall i~\cdot$ 
\item[] c) \ $G_o(\vec{r}, \vec{r} \, ') = 0$ at $\vec{r} \in S~\cdot$ 
\end{itemize}

\noindent If we integrate condition (a) \ then consider the spherical symmetry $G(\vec{r}, \vec{r} \ ') = G(| \vec{r} - \vec{r} \, ' |) = G(\scripty{r})$ with the fact that \textbf{\^ n} is parallel to $\vec{r} - \vec{r} \, '$, then:

\begin{equation} \label{eq.1.5}
\int \limits_S \nabla G(\vec{r}, \vec{r} \ ') . \textbf{\^ n} \ dS = 2 \pi \scripty{r} \ \frac{\partial G(\scripty{r})}{\partial \scripty{r}} \Biggr|_{\scripty{r}=R} = 1 \, .
\end{equation}
When $G_o(\scripty{r}) \to 0$ as $|\scripty{r} \ | \to \infty$, then Sommerfeld got the \textit{logarithmic potential}, Neumann had studied before, for the upper surface as:

\begin{equation} \label{eq.1.6} 
G(\vec{r}, \vec{r}_{up}^{ \ '}) = \frac{1}{2 \pi} \ln{| \vec{r} - \vec{r}_{up}^{ \ '} | + C}~,
\end{equation}
where $C$ is determined according to the boundary condition. If we want Green function to vanish at $\vec{r} \to \infty$ then $C=0$.\\

In Sommerfeld treatment there is only one copy of double sheeted Riemann surface. Then we expect one mirror image function in the lower surface to be similar to that of the upper surface, up to sign. And the general Green function is
 
\begin{equation} \label{eq.1.7}
G_o(\vec{r}, \vec{r} \ ') = \frac{1}{2 \pi} \ln{| \vec{r} - \vec{r}_{up}^{ \ '} |} - \frac{1}{2 \pi} \ln{| \vec{r} - \vec{r}_{down}^{ \ ''} |} ~\cdot
\end{equation}
Finally by combining \eqref{eq.1.7} and \eqref{eq.1.3}, the solution of potential function $u(\vec{r})$ is obtained.

\chapter{2D Grounded Circular Ring}
\hspace{\parindent} Wormhole studies have captured interests since the introduction of the \textit{Einstein-Rosen bridge} \cite{PPG}. The classic Ellis wormhole \cite{Eft} can be seen in 2D, after been foliated and sliced, as circles with redii $r(w) = \sqrt{R^2 + w^2}$ where $w \in (- \infty , + \infty)$ is the parameter that relates the coordinates of vertical axis $z$, with circles aligned around it, with coordinates of curves $r$ that describes how to move from circle to another. $R$ is the fixed radius of the smallest circle at the equatorial near the bridge that connects the double sheets of Riemann surface.\\
	Our approach is based on building Laplace-Beltrami operator of such manifold to get both the fundamental solution and the image solution of Green functions after considering the required Dirichlet boundary conditions together with the corresponding discontinuities.\\
	Next we strict the radii of those aligned circles to become $r(w) = (R^p + |w|^p)^{1/p}$ for $p \in [1,2]$ and see how Green functions behave upon taking the limit $p \to 1$ where both branch of wormhole become completely \textquotedblleft squashed flat\textquotedblright. The case $r(w) = R + |w|$ is the Manhatten norm and equatorial circle with radius $R$ serves as a \textquote{portal} or a \textquote{doorway}. Following Sommerfeld's method, the interiors of these circles act as forbidden region of Sommerfeld screen; they are excluded from Riemann surfaces too.

	\section{Logarithmic potential}
\hspace{\parindent}	For a regular grounded conducting ring with radius $R$ in Euclidean 2D space $\mathbb{E}$, the electrostatic potential function of a point-like charge at $\vec{r} \ '$ is given by:
	\begin{equation} \label{eq.2.1}
	\begin{split}	
	\Phi_{\mathbb{E}_2} (\vec{r}-\vec{r} \ ') = - \frac{1}{2\pi} \ln{\left( \frac{|\vec{r}-\vec{r} \ '|}{R} \right)} \ , \qquad \qquad
	\nabla^2 \Phi_{\mathbb{E}_2} (\vec{r}-\vec{r} \ ') = - \delta^2 (\vec{r}-\vec{r} \ ') \, .
	\end{split}
\end{equation}	 
And the corresponding Green function obeying $\nabla^2 G(\vec{r}, \vec{r} \ ')=-\delta^2(\vec{r}-\vec{r} \ ')$ is:
\begin{equation} \label{eq.2.2}
\begin{split}
G(\vec{r},\vec{r} \ ') & = - \frac{1}{2\pi} \ln{\left( \frac{|\vec{r}-\vec{r} \ '|}{R} \right)} \\
& = - \frac{1}{2\pi} \left[ \ln{\left( \frac{r_>}{R} \right) + \ln{\left( 1+ \frac{r_<^2}{r_>^2} - 2\frac{r_<}{r_>} \cos(\theta) \right)}} \right] ,
\end{split}
\end{equation}
 where $r_\gtrless = \stackanchor{max}{min} \ (\vec{r},\vec{r} \ ')$ and $\theta$ is the angle between $\vec{r}$ and $\vec{r} \ '$.\\
 For the image inside the ring at position $\displaystyle{\vec{r} \ '' = \frac{R^2}{r'^2} \vec{r} \ '}$ Green function is:
  
\begin{equation} \label{eq.2.3}
 G_1(\vec{r},\vec{r} \ '') = \frac{1}{2\pi} \ln{\left( \frac{|\vec{r}-\vec{r} \ ''|}{R} \right)} \, .
\end{equation}  
Then the general Green function is:
\begin{equation} \label{eq.2.4}
G_o(\vec{r},\vec{r} \ ') = - \frac{1}{2\pi} \ln{\left( \frac{|\vec{r}-\vec{r} \ '|}{R} \right)} + \frac{1}{2\pi} \ln{\left( \frac{|\vec{r}-\frac{R^2}{r'^2} \vec{r} \ '|}{R} \right)} + \frac{1}{2\pi} \ln{\left( \frac{|\vec{r} \ '|}{R} \right)} \, .
\end{equation}
The last term is some constant $C$ we referred to before in \eqref{eq.1.6} that's to be determined according to the b.c's when $\displaystyle{G_o(\vec{r},\vec{r} \ ') \vert_{r' = R}} = 0$. The last two terms are harmonic for all $\vec{r}, \vec{r} \ '$, i.e., they obey the condition $\nabla^2 G_i(\vec{r}, \vec{r_i} \, ' )=0$ we stated before, and $G(\vec{r},\vec{r} \ ')$ is the only one contributes the Dirac delta since it's the fundamental Green function. Those additional image functions break translational invariance on the plane but, of course, they don't change the symmetry of the arguments of the Green function upon changing $\vec{r} \leftrightarrow \vec{r} \ '$.
\section{Green functions for 2D curved surface} \label{sec.2.2}

\hspace{\parindent} A 2D Riemannian manifold a topological space that is endowed with a metric $g_{\mu \nu}$, where $g \equiv$ det $g_{\mu \nu}$ and $1 \leq \mu, \nu \leq 2$, with space intervals $ds^2 = g_{\mu \nu} dx^{\mu}dx^{\nu}$. Then the invariant Laplace-Beltrami operator $ \nabla^2 = \frac{1}{\sqrt{g}} \partial_{\mu} (\sqrt{g} g^{\mu \nu} \partial_{\nu})$. For such manifold, the space interval is:
\begin{equation} \label{eq.2.5}
(ds)^2=(dw)^2+r^2(w)\ (d\theta)^2 \, ,
\end{equation}
and its invariant Laplacian\footnote{A reader might wonder why to express it in partial derivatives rather than $\nabla$ covariant ones. The \textit{Christoffel connections} are implicitly expressed as $\Gamma_{\mu \nu}^{\nu} = \frac{1}{\sqrt{g}} \partial_{\mu}(\sqrt{g})=\frac{1}{2} g^{\alpha \beta}\partial_{\mu}(g_{\alpha \beta})$ upon applying Leibntz rule to the Laplacian components. Remember the operator as whole is acting on a scalar function.} is:
\begin{equation} \label{eq.2.6}
\nabla^2 = \frac{1}{r(w)} \left[\partial_w(r(w)\partial_w)+\partial_{\theta}^2\right].
\end{equation} 
Upon changing variable:
\begin{equation} \label{eq.2.7}
u(w)= \int \limits_0^w \frac{d\tilde{w}}{r(\tilde{w})} \ ,
\end{equation}
the Laplacian can be read more clearly as:
\begin{equation} \label{eq.2.8}
\nabla^2 = \frac{1}{g} \left( \frac{\partial^2}{\partial u^2} + \frac{\partial^2}{\partial \theta^2} \right) \, .
\end{equation}
Now harmonic functions $h$  corresponding to Laplace equation $\nabla^2 h = 0$ are periodic for fixed $w$ in $\theta \in [0,2\pi]$. Therefore they are:
\begin{equation} \label{eq.2.9}
h_0 (u)=a+b \ u \quad \textnormal{and} \quad h_m^{\pm}(u,\theta)=c_m \ e^{-mu} \ e^{\pm i m \theta} \quad \textnormal{for} \ m \in \mathbb{Z} \textnormal{\textbackslash} \{0\} \, .
\end{equation}
After change of variables \eqref{eq.2.9} can be written as:
\begin{equation} \label{eq.2.10}
h_0 (w)=a+b \int \limits_0^w \frac{d\tilde{w}}{r(\tilde{w})} \quad \textnormal{and} \quad h_m^{\pm}(w,\theta)=c_m \ e^{\pm i m \theta} \ \exp \left( -m\int \limits_0^w \frac{d\tilde{w}}{r(\tilde{w})} \right) \, .
\end{equation}
Notice that $h_0$ is isotropic, so for $w_{\gtrless} = \stackanchor{max}{min}(w,w')$ the radial part of Laplace equation for $h_0(u)$ can be read as:
\begin{equation} \label{eq.2.11}
\begin{split}
\frac{d}{dw} \left[ r(w) \frac{d}{dw} \left( \int_{w_<}^{w_>} \frac{d\tilde{w}}{r(\tilde{w})} \right) \right] & =  \frac{d}{dw} \bigg\{ \substack{+1 \ , \ w>w' \\ \\ \\  \\ -1 \ , \ w<w'} \qquad \Longrightarrow \\ \frac{d}{dw} \left[ \Theta(w-w') - \Theta(w'-w)\right] & = 2 \delta(w-w') \, ,
\end{split}
\end{equation}
where $$\Theta(w-w') = \Big\{ \substack{+1 \ , \ w>w' \\ \\  \\ 0 \ , \ w<w'}$$ is the Heaviside function that arises from discontinuities in radial derivatives of such harmonic functions as in \eqref{eq.1.2}. Such discontinuity would determine $b$ as:
\begin{align}\label{eq.1.12}
\begin{split}
-\frac{1}{4\pi} \lim \limits_{\varepsilon \to 0} \int \limits_{w'-\varepsilon}^{w'+\varepsilon} \frac{d}{dw} \left[ r(w) \frac{d}{dw} \left(~\int\limits_{w_<}^{w_>} \frac{d\tilde{w}}{r(\tilde{w})} \right) \right]dw & = -\frac{1}{4\pi}\lim \limits_{\varepsilon \to 0} \left[ r(w) \frac{d}{dw} \left(~\int \limits_{w_<}^{w_>} \frac{d\tilde{w}}{r(\tilde{w})} \right) \right]_{w=w'-\varepsilon}^{w=w'+\varepsilon} \\
& = -\frac{1}{2\pi}~,
\end{split}
\end{align}
hence $b=-1/(4\pi)$.\\
Similarly $\nabla^2 h_m(u,\theta)=0$ gives another discontinuity\footnote{By choosing Green function as in \eqref{eq.2.18} we avoid the $\sin(m(\theta-\theta'))$ term that ought to contribute to the $\delta(\theta-\theta')$ in \eqref{eq.2.17}. Equation \eqref{eq.2.15} proves the consistency of not having such sine term.}:
\begin{equation} \label{eq.2.13}
\frac{d}{dw} \left\lbrace r(w) \frac{d}{dw} \left[ \exp \left( -m \int_{w_<}^{w_>} \frac{d\tilde{w}}{r(\tilde{w})} \right) \right] \right\rbrace =
-2m \ \delta(w-w') \ + \ m^2 \exp \left( -m \int_{w_<}^{w_>} \frac{d\tilde{w}}{r(\tilde{w})} \right) \, .
\end{equation}
Next we solve the coefficients of \eqref{eq.2.9} $c_m=1/(2\pi m)$ if we  divide the last equation by $m$ and use $\displaystyle{\sum \limits_{m=-\infty}^{\infty} e^{im(\theta-\theta')} = 2\pi \ \delta(\theta-\theta')}$ together with:
\begin{equation} \label{eq.2.14}
\begin{split}
\frac{1}{2\pi}\lim \limits_{\varepsilon \to 0} \int \limits_{w'-\varepsilon}^{w'+\varepsilon} \frac{d}{dw} \left\lbrace r(w) \frac{d}{dw} \left[ \frac{1}{m} \left( e^{-m \int_{w_<}^{w_>} \frac{d\tilde{w}}{r(\tilde{w})}} \right) \cos(m(\theta-\theta'))\right] \right\rbrace dw = \frac{- \cos{(m(\theta-\theta'))}}{\pi}~.
\end{split}
\end{equation}
Then using $\displaystyle{\sum \limits_{n=1}^{\infty} \frac{z^n}{n} = -\ln(1-z)}$ we can expand $h_m(u,\theta)$ as following:
\begin{equation} \label{eq.2.15}
\begin{split}
\sum \limits_{m=1}^{\infty} h_m^{\pm}(u,\theta)-h_m^{\pm}(u',\theta') & = \frac{1}{2\pi} \sum \limits_{m=1}^{\infty} \frac{1}{m} \ \left[ \ e^{-mu} e^{-im\theta} - e^{-mu'} e^{im\theta'} \ \right] \\
&~\\
& = - \frac{1}{4\pi}\left[ \ \ln(1-e^{-mu} e^{-im\theta}) + \ln(1-e^{-mu'} e^{im\theta'}) \ \right]\\
&~\\
& = - \frac{1}{4\pi} \ \ln \left[ \ 1-(e^{-m(u+u')})^2 -2e^{-m(u+u')}\cos{(m(\theta-\theta'))} \ \right] \, .
\end{split}
\end{equation}
For $z < 1$ , the last equation is in form of $\ln{[1+z^2-2z\cos(m(\theta-\theta'))]}$ which,  using Taylor series, can be expanded then summed into:
\begin{equation} \label{eq.2.16}
\begin{split}
\ln\left[1+(e^{-m(u+u')})^2-2 e^{-m(u+u')} \cos(m(\theta-\theta'))\right] = -2 \sum \limits_{m=1}^{\infty} \frac{(e^{-m(u+u')})^2}{m} \cos{\left(m(\theta-\theta')\right)} \, .
\end{split}
\end{equation}
Finally \textquote{a} fundamental Green function obeying:
\begin{equation} \label{eq.2.17}
\nabla^2 G(w,\theta;w',\theta')=-\frac{1}{\sqrt{g}} \ \delta(w-w')  \ \delta(\theta-\theta') \, ,
\end{equation}
is given by:
\begin{equation} \label{eq.2.18}
\begin{split}
G(w(u),\theta;w'(u'),\theta') & = [h_0(u(w))-h_0(u'(w'))]+ [h_m^{\pm}(u,\theta)-h_m^{\pm}(u',\theta')] \\
&~\\
& = - \frac{1}{4\pi} \left[ \int\limits_{0}^{w_>} \frac{d\tilde{w}}{r(\tilde{w})} - \int\limits_{0}^{w_<} \frac{d\tilde{w}}{r(\tilde{w})} \right] + \sum \limits_{m=1}^{\infty} c_m \ e^{m(u(w)+u'(w'))} \ e^{im(\theta-\theta')} \\
&~\\
& = - \frac{1}{4\pi} \int\limits_{w_<}^{w_>} \frac{d\tilde{w}}{r(\tilde{w})} + \frac{1}{2\pi} \sum \limits_{m=1}^\infty \frac{1}{m} \ e^{ -m \int_{w_<}^{w_>} \frac{d\tilde{w}}{r(\tilde{w})}} \ \cos(m(\theta-\theta'))\\
&~\\
& = - \frac{1}{4\pi} \int\limits_{w_<}^{w_>} \frac{1}{r(\tilde{w})}d\tilde{w} \\
&~\\
& \quad - \frac{1}{4\pi} \ln \Big[ 1+e^{-2(\int_{w_<}^{w_>} \frac{d\tilde{w}}{r(\tilde{w})})} -2e^{-(\int_{w_<}^{w_>} \frac{d\tilde{w}}{r(\tilde{w})})} \cos{(m(\theta-\theta'))} \Big].
\end{split}
\end{equation}\\
We incorporate vanishing b.c's at $w=\pm \infty$ for the non-isotropic terms by assuming $\displaystyle{\int_{w_<}^{w_>} \frac{d\tilde{w}}{r(\tilde{w})}}$ diverges as $w_{\gtrless} \to \pm \infty$, leaving only the isotropic term in Green function. Moreover, translational invariance of such Green function is lost for general $r(w)$. However, For such wormhole $r(w)=(R^{1/p}+|w|^{1/p})^{1/p}$ \ it can be regained only when the equatorial ring that connects two upper and lower branches, shrinks to a point, i.e., $r(w)=|w|$.\\
To finish constructing Green function we need to consider the case where $w'=0$, i.e., that corresponds to grounding the equator of the throat. So for a surface symmetrical under $w \leftrightarrow -w$ we need to add the mirror image Green function $G(w,\theta;w'',\theta'')$ which is contributed from the negative Kelvin image \footnote{To a distant observer on the source charge branch the image charge \textit{appears} to be located inside the equatorial circle of the wormhole's throat due to \textit{lensing} effects.} on the other side of the throat, i.e., $w''=-w'$, and $\theta'=\theta''$ due to rotational invariance. If the manifold is $p$-norm wormhole, then more general Green function that describes electrostatic potential function everywhere on both upper and lower branches for such wormhole with grounded equator ($ring$  of radius $R$) in the middle of the manifold bridge is:
\begin{equation}\label{eq.2.19}
G_o(w,\theta;w',\theta')=G(w,\theta;w',\theta')-G(w,\theta;-w',\theta')~,
\end{equation}
which is what we discuss next chapter.
 
\section{$p$-norm and Ellis wormholes}

\hspace{\parindent} Before we study Green functions on an 2D wormhole background defined by:
\begin{equation}\label{eq.2.20}
r(w)=(R^{p}+|w|^{p})^{1/p}, \qquad \text{for} \ p \in [1,2]~,
\end{equation}
and is embedded in 3D space as shown in \cite{Crw} Appendix.A, we need to check curvature singularities of such manifold, i.e., we need to get the Gaussian curvature by calculating the scalar curvature from Riamann curvature tensor \cite{Kdg}. \\
We describe the extrinsic geometry, or the embedding in $\mathbb{E}_3$, of 2D surfaces in terms of the normal component of rate change of local coordinate basis ($\mathbf{\hat{e}}_{\nu}=\partial_{\mu} \vec{\mathbf{r}})$ of the coordinate system. This is called the $2^{\underline{nd}}$ \textit{fundamental form} (the \textit{shape operator} or Weingarten equation):
\begin{equation}\label{eq.2.21}
\mathbf{\Pi}_{\mu \nu}= -\mathbf{\hat{e}}_{\nu}. \nabla_{\mu} \textbf{\^ n} = \nabla_{\mu} \mathbf{\hat{e}}_{\nu}. \textbf{\^ n} = \Gamma_{\mu \nu}^{\rho} \mathbf{\hat{e}}_{\rho}. \textbf{\^ n} \, ,
\end{equation}
where: 
\begin{equation}\label{eq.2.22}
\Gamma_{\rho \nu}^{\mu} = \frac{1}{2} g^{\mu \sigma} (\partial_{\rho}g_{\sigma \nu}+\partial_{\nu}g_{\sigma \rho}-\partial_{\sigma}g_{\nu \rho}) \, ,
\end{equation}
 is the \textit{Christoffel connection} of the $2^{\underline{nd}}$ kind.\\
\textit{Theorema Egregium} asserts that although we can express the Gaussian curvature of a 2D surface as a ratio between the determinants  of both $1^{\underline{st}}$ and $2^{\underline{nd}}$ \textit{fundamental forms,} $(\mathbf{I})$ and $(\mathbf{\Pi})$ respectively, i.e., $\displaystyle{\mathcal{K}=\frac{\det((\mathbf{\Pi)}}{\det(\mathbf{I})}}$, the curvature is still dependent only on the $1^{\underline{st}}$ fundamental form and its derivatives, i.e., it's intrinsic property.\\
This can be shown if we combine \textit{Codazzi-Mainardi} equation:
\begin{equation}\label{eq.2.23}
\partial_{\rho} \mathbf{\Pi}_{\mu \nu} - \partial_{\nu} \mathbf{\Pi}_{\mu \rho} = \Gamma_{\mu \rho}^{\lambda} \mathbf{\Pi}_{\lambda \nu} - \Gamma_{\mu \nu}^{\lambda} \mathbf{\Pi}_{\lambda \rho} \, ,
\end{equation}
together with the Riemann curvature tensor:
\begin{equation} \label{eq.2.24}
R_{ \ \nu \rho \sigma}^{\mu} = \partial_{\rho} \Gamma_{\nu \sigma}^{\mu} - \partial_{\sigma} \Gamma_{\nu \rho}^{\mu} + \Gamma_{\rho \lambda}^{\mu} \Gamma_{\nu \sigma}^{\lambda} - \Gamma_{\sigma \lambda}^{\mu} \Gamma_{\nu \rho}^{\lambda}  \, ,
\end{equation}
and  the scalar curvature we get from the Ricci curvature tensor as follows:
\begin{equation}\label{eq.2.25}
R_{\mu \nu}=R_{ \ \mu \rho \nu}^{\rho} \, ,
\end{equation}
\begin{equation}\label{eq.2.26}
\mathcal{R}=R_{ \ \mu}^{\mu}=g^{\mu \nu} R_{\mu \nu} \, .
\end{equation}
Notice in $n$D manifolds the Riemann tensor has $n^4$ components. However, due to its skew-symmetric properties, then in 2D the Riemann tensor reduced to only $R_{ \ 212}^{1}$ and $R_{ \ 121}^{2}=-R_{ \ 112}^{2}=-(-R_{ \ 212}^{1})$.
Using the previous results [\eqref{eq.2.21}-\eqref{eq.2.26}] \textit{Theorema Egregium} reveals that for 2D, and only 2D, surfaces:
\begin{equation}\label{eq.2.27}
\mathcal{K}=\frac{1}{2}\mathcal{R}=\frac{R_{1212}}{\det(g)}~ \cdot
\end{equation}
Now to calculate the curvature of $p$-norm wormhole we have $R_{ \ 212}^{1}$ in form of:
\begin{equation}\label{eq.2.28}
R_{ \ \theta w \theta}^{w} = \partial_{w} \Gamma_{\theta \theta}^{w} - \partial_{\theta} \Gamma_{\theta w}^{w} + \Gamma_{w \lambda}^{w} \Gamma_{\theta \theta}^{\lambda} - \Gamma_{\theta \lambda}^{w} \Gamma_{\theta w}^{\lambda} \, ,
\end{equation}
where we sum over $\lambda \in \{w, \theta \}$.\\
Then using \eqref{eq.2.22} we have:
\begin{equation} \label{eq.2.29}
\begin{split}
\Gamma_{\theta \theta}^{w} & = \frac{1}{2} \partial_w [r(w)]^2  \, , \\
\Gamma_{\theta w}^{\theta} & = \frac{1}{2} \ \frac{1}{[r(w)]^2} \ \partial_w [r(w)]^2  \, ,\\
\Gamma_{\theta \theta}^{\theta} & = \Gamma_{w w}^{w} = \Gamma_{w \theta}^{w} = \Gamma_{w w}^{\theta} = 0 ~ .
\end{split}
\end{equation}
And finally \eqref{eq.2.27} becomes:
\begin{equation}\label{eq.2.30}
\begin{split}
\mathcal{K} = \frac{ g_{ww} R_{ \ \theta w \theta}^{w}}{\det (g)} & = \frac{ g_{ww}(\partial_{w} \Gamma_{\theta \theta}^{w} - \Gamma_{\theta \theta}^{w} \Gamma_{\theta w}^{\theta})} {\det (g)} \\
& = \frac{1}{[r(w)]^2} \left\lbrace \frac{1}{2} \  \partial_w^2 \bigg[ r(w) \bigg]^2- \bigg[ \frac{1}{2}\frac{1}{r(w)}\partial_w(r(w))^2 \bigg]^2 \right\rbrace \\
& = - \frac{1}{r(w)} \ \frac{d^2r(w)}{dw^2}\\
& = \frac{(1-p) \ R^2 \ (w^2)^{(p-2)/2}}{\big(R^p+(w^2)^{p/2}\big)^2} ~\cdot
\end{split}
\end{equation}
The last result for the curvature assures that for $p=1$ i.e., the Manhattan norm, the manifold is flat so the upper and the lower branches are perfectly superimposed over each other. Moreover, when we get close to the equatorial ring, i.e., $w \to 0$, the radius corresponding to such curvature blows up $\forall p \in [1,2]$. Then the equatorial ring (the \textit{bridge}) becomes made of points that all are singular. More details on that are in \cite{Crw}, Appendix C.\\  
Final comment of the geometry before we return back to Green functions business. We may notice:
\begin{equation}\label{eq.2.31}
\begin{split}
|w|& =r\left(1-(\frac{R}{r})^p\right)^{1/p}\\
& = r \ \left(1-\frac{1}{p} \ \frac{R^p}{R^p+|w|^p}+\cdots\right) \, .
\end{split}
\end{equation}
So when $w \to \pm \infty$ , the distance interval becomes:
\begin{equation}\label{eq.2.32}
(ds)^2 \substack{\\ \\ \approx \\ w \to \pm \infty} \ (dr)^2+r^2(d\theta)^2,
\end{equation}
 which guarantees such wormholes own flat curvature when $w \to \pm \infty$. It's obvious when in the case of Manhattan norm for $r \gg R$.\\
 
Now for Green functions of different $p$-norms, there is a special case where $p=2$ (the \textit{Ellis wormhole}) as \eqref{eq.2.17} becomes:
\begin{equation}\label{eq.2.33}
\begin{split}
\nabla^2 G(w,\theta;w',\theta')& = \frac{1}{\sqrt{R^2+w^2}} \ \partial_w \left( \sqrt{R^2+w^2} \ \partial_w G \right) \ + \ \frac{1}{R^2+w^2} \ \partial_{\theta}^2 G \\
& = -\frac{1}{\sqrt{g}} \ \delta(w-w')  \ \delta(\theta-\theta') ~ ,
\end{split}
\end{equation} 
 while \eqref{eq.2.7} becomes:
\begin{equation} \label{eq.2.34}
\begin{split}
-\frac{1}{4\pi R} \ \int \limits_{w_<}^{w_>} \frac{d\tilde{w}}{\sqrt{1+ \frac{\tilde{w}^2}{R^2}}} & = -\frac{1}{4\pi R} \ln\left[\sqrt{\frac{\tilde{w}^2}{R^2}+1} + \frac{\tilde{w}}{R} \right]_{w_<}^{w_>}\\
& = \frac{1}{4\pi} \ln \left[ \left( \sqrt{\frac{w_>^2}{R^2}+1} - \frac{w_>}{R} \right) \ \left( \sqrt{\frac{w_<^2}{R^2}+1} + \frac{w_<}{R} \right) \right] \, .
\end{split}
\end{equation}
We prefer expressing the solution of that integral as logarithmic function rather than hyperbolic sine function as it's related to the Green function of the logarithmic potential.
Substitute \eqref{eq.2.34} in \eqref{eq.2.18} to get:
\begin{equation}\label{eq.2.35}
\begin{split}
G(w,\theta;w',\theta') = & + \frac{1}{4\pi} \ln \left[ \left( \sqrt{\frac{w_>^2}{R^2}+1} - \frac{w_>}{R} \right) \left( \sqrt{\frac{w_<^2}{R^2}+1} + \frac{w_<}{R} \right) \right]\\
& - \frac{1}{4\pi} \ln \left[1 + \left( \sqrt{\frac{w_>^2}{R^2}+1} - \frac{w_>}{R} \right)^2 \left( \sqrt{\frac{w_<^2}{R^2}+1} + \frac{w_<}{R} \right)^2 \right] \\ 
& -\frac{1}{4\pi} \left[-2\left( \sqrt{\frac{w_>^2}{R^2}+1} - \frac{w_>}{R} \right) \left( \sqrt{\frac{w_<^2}{R^2}+1} + \frac{w_<}{R} \right) \cos(\theta-\theta') \right].
\end{split}
\end{equation}
Then substitute it in \eqref{eq.2.19} to get the desired general Green function $G_o(w,\theta;w',\theta')$. Now we are ready to squash Ellis wormholes!

\section{Squashing $p$-norm wormholes into Manhattan wormhole} \label{sec.2.4}

\hspace{\parindent} Wormhole \textquote{flattening} is achieved as a continuous deformation by taking the radial function to be the $p$-norm of $R$ and $w$. That is, a $\pm w$ symmetric p-norm wormhole in 2D is defined taking $p=1$ \footnote{\label{ftnt6}In the opposite extreme, where $p \to \infty$, the $p$-norm wormhole function becomes the right-circular cylinder of \cite{Viw}, p.488, eqn(3).} so the throat \textquote{height} becomes zero. That is:
\begin{equation}\label{eq.2.36}
r(w)=R+|w| \, ,
\end{equation}
Then for the case where both $w$ and $w'$ are on the same side from the branch cut of the double-sheeted space, i.e., the same wormhole branch, the integral \eqref{eq.2.7} is:
\begin{equation}\label{eq.2.37}
\int \limits_{w_<}^{w_>} \frac{d\tilde{w}}{R+|\tilde{w}|}=\ln\left(\frac{R+w_>}{R+w_<}\right).
\end{equation}
Meanwhile for the other case where $w$ and $w'$ are on opposite side from the branch cut of the double-sheeted space, let's say $w > 0$ and $w' < 0$, the integral \eqref{eq.2.7} is:
\begin{equation}\label{eq.2.38}
\int \limits_{w_<}^{w_>} \frac{d\tilde{w}}{R+|\tilde{w}|}=\ln\left(\frac{R+w}{R}\right)+\ln\left(\frac{R-w'}{R}\right) \, .
\end{equation}\\
For the first scenario, \eqref{eq.2.18} becomes:
\begin{equation}\label{eq.2.39}
\begin{split}
G(w,\theta;w',\theta') \vert_{\substack{\\ w>0\\w'>0}} & =
- \frac{1}{4\pi} \ln \Bigg[ \left[\frac{R+w_>}{R+w_<}\right]\left[1+\left[\frac{R+w_>}{R+w_<}\right]^2 - 2 \left[\frac{R+w_>}{R+w_<}\right] \cos{(\theta-\theta')}\right] \Bigg]\\
& = - \frac{1}{4\pi} \ln\left(\frac{r^2+r'^2-2rr'\cos{(\theta-\theta')}}{rr'}\right)~.
\end{split}
\end{equation}
While the second scenario makes \eqref{eq.2.18}to be:
\begin{equation}\label{eq.2.40}
G(w,\theta;w',\theta') \vert_{\substack{w>0\\w'<0}} = -\frac{1}{4\pi} \ln \left(\frac{rr'}{R^2}+\frac{R^2}{rr'}-2\cos{(\theta-\theta')}\right).
\end{equation}
where $r=R+|w|$ and $r'=R+|w'|$.
The last two equations \eqref{eq.2.39} and \eqref{eq.2.40} describe the general Green function for any $w,w'$. We may check the consistency of these two equations with the conventional Green functions got by Kelvin's method for \textquote{the} traditional single Euclidean plane at identical points. After neglecting the azimuthal dependence, the difference for the first case is given by:
\begin{equation}\label{eq.2.41}
G(w,\theta;w',\theta') \vert_{\substack{w>0\\w'>0}} - G_{\mathbb{E_{\text{2}}}}(\vec{r},\vec{r} \ ') = -\frac{1}{4\pi} \ln\left(\frac{R^2}{rr'}\right),
\end{equation}
while the difference in the second case is:
\begin{equation}\label{eq.2.42}
G(w,\theta;w',\theta') \vert_{\substack{w>0\\w'<0}} - G_{\mathbb{E_{\text{2}}}}(\vec{r},\frac{R^2}{(\vec{r} \ ')^2} \ \vec{r} \ ') = \frac{1}{4\pi} \ln\left(\frac{r}{r'}\right),
\end{equation}
where Kelvin's method restricts the image point inside the ring (the equatorial circle in Sommerfeld's method) at $\mathfrak{\vec{r} \ '}=\frac{R^2}{(\vec{r} \ ')^2} \ \vec{r} \ '$. Since in Kelvin's method $r'\gneq R$, then $\mathfrak{\vec{r} \ '}=\frac{R^2}{(\vec{r} \ ')^2} \ \vec{r} \ \lneq R'$. And obviously the RHS's of \eqref{eq.2.41} and \eqref{eq.2.42} are harmonic on $\mathbb{E_{\text{2}}}$ where $r \neq 0 \neq r'$.\\
If we substitute \eqref{eq.2.39} in \eqref{eq.2.19}, the more general Green function that describes electrostatic potential function everywhere on both upper and lower branches for Manhattan wormhole with grounded equator ($ring$  of radius $R$) in the middle of the manifold bridge becomes:
\begin{eqnarray}\label{eq.2.43}
G_o(w,\theta;w',\theta') \vert_{\substack{w \gtrless 0 \\ w' \gtrless 0}} \!\!\!& =&\!\!\! -\frac{1}{4\pi}\ln\left[\frac{r^2+r'^2-2rr'\cos{(\theta-\theta')}}{rr'}\right]\nonumber\\
\!\!\!&&\!\!\! +\frac{1}{4\pi}\ln\left[\frac{rr'}{R^2}+\frac{R^2}{rr'}-2\cos{(\theta-\theta')}\right]\nonumber\\
\!\!\!& =&\!\!\! -\frac{1}{4\pi}\ln\left[R^2\frac{r^2+r'^2-2rr'\cos{(\theta-\theta')}}{r^2r'^2+R^4-2R^2rr'\cos{(\theta-\theta')}}\right] \, . 
\end{eqnarray}
If we substitute \eqref{eq.2.40} in \eqref{eq.2.19} then $G_o(w,\theta;w',\theta') \vert_{\substack{w \gtrless 0 \\ \\ w' \lessgtr 0}} = - G_o(w,\theta;w',\theta') \vert_{\substack{w \gtrless 0 \\ \\ w' \gtrless 0}}$ as it's an odd function of $w$ for fixed $w'$.\\
The corresponding general Green function, for both the source charge and its image, obtained by Kelvin's method is given by:
\begin{equation}\label{eq.2.44}
G_{o \ \mathbb{E_{\text{2}}}}(\vec{r},\vec{r} \ ') = G_{\mathbb{E_{\text{2}}}}(\vec{r},\vec{r} \ ')-G_{\mathbb{E_{\text{2}}}}(\vec{r},\frac{R^2}{(\vec{r} \ ')^2} \ \vec{r} \ ') \ .
\end{equation}
Then there is \textit{disparity} shown in the difference between the general Green function anywhere on the squashed wormhole and the general Green function of Kelvin's method:
\begin{equation}\label{eq.2.45}
G_o(w,\theta;w',\theta') - G_{o \ \mathbb{E_{\text{2}}}}(\vec{r},\vec{r} \ ') = - \frac{1}{4\pi} \ln \left(\frac{R}{r'}\right) \, .
\end{equation}
We notice that such disparity comes from the fact that $G_{o \ \mathbb{E_{\text{2}}}}$ is \emph{not} symmetric under $\vec{r}\leftrightarrow \vec{r} \ '$. However since the difference is harmonic on both the Euclidean plane and the Riemannian wormhole for any $r'\neq 0$, then both $\nabla^2 \bigg|_{r,r'} G(\vec{r},\vec{r} \ ') = - \delta(\vec{r},\vec{r} \ ')$ and Dirichlet condition $G(\vec{r},\vec{r} \ ') \big|_{\vec{r} \ ' = R \text{\^{r}} } =0$ are still satisfied for any points outside the forbidden region inside the Euclidean ring corresponding to the Riemannian \textit{bridge} regardless whether Green function is considered to be $G_o$ or $G_{o \ \mathbb{E_{\text{2}}}}$ .

\section{Relating Kelvin and Sommerfeld images \\ through inversion transformation} \label{sec.2.5}
\hspace{\parindent} Part of complex structure that is endowed to Riemann surfaces is inversion mapping:
\begin{equation}\label{eq.2.46}
\mathfrak{I} : \vec{r} \mapsto \vec{\mathfrak{r}} = \frac{R^2}{r^2} \vec{r} ~ ,
\end{equation}
where $R$ is the radius of Riemann's sphere.\\ Due to angular symmetry $\mathfrak{\hat{r}}$ = \textbf{\^{r}}, we look to the traditional ring, with same radius $R$, as an $S^1$ bounding Riemann sphere. We can relate the mirror image got by Kelvin's method to that image got by Sommerfeld's method through inversion transformation.\\
Under such transformation Laplace-Beltrami operator is transformed as:
\begin{equation}\label{eq.2.47}
\mathfrak{I}(\nabla_r^2) = \left(\frac{R^2}{r^2}\right)^2 \left(\nabla_{\mathfrak{r}}^2 \  + \ \frac{2(2-N)}{\mathfrak{r}^2}D_{\mathfrak{r}}\right), \qquad \qquad D_{\mathfrak{r}} = \vec{\mathfrak{r}} \cdot \vec{\nabla_{\mathfrak{r}}} ~ ,
\end{equation}
such that, \textit{except} for 2D, the Laplacian \cite{Ghw} is not an eigenvector of such transformation, it has extra piece (\textit{scale operator}) to be added to it under inversion. In 2D case, Green functions are invariant under inversion as:
\begin{equation}\label{eq.2.48}
\begin{split}
\mathfrak{I} \left( \, G(\vec{r},\vec{r} \ ') \, \right) & =\frac{(\vec{\mathfrak{r}})^{N-2} \, (\vec{\mathfrak{r}} \, ')^{N-2}}{R^{2N-4}} \, G(\vec{\mathfrak{r}},\vec{\mathfrak{r}} \ ')\\
\\
& = G(\vec{\mathfrak{r}},\vec{\mathfrak{r}} \ ')~ , \qquad \qquad \text{for} \ N=2~\cdot
\end{split}
\end{equation}
And:
\begin{equation}\label{eq.2.49}
\begin{split}
\mathfrak{I} \big[\nabla_r^2 G(r,r')\big] & = \mathfrak{I} \big[-\frac{1}{\sqrt{g(r)}} \delta(r-r') \delta(\theta-\theta')\big] \\
& \downarrow\\
\nabla_{\mathfrak{r}}^2 G(\mathfrak{r},\mathfrak{r}') & = -\frac{1}{\sqrt{g(\mathfrak{r})}} \delta(\mathfrak{r}-\mathfrak{r}') \delta(\theta-\theta') \, ,
\end{split}
\end{equation}
leaving also the condition on Green functions unchanged under inversion.\\

So these results lead us to map the lower branch of Manhattan norm wormhole to the interior of the Euclidean ring, while it leaves the upper branch unchanged. The more general case for $N\neq 2$ is to be discussed in next chapter.

\chapter{$n$D Grounded Conducting Hyperspheres}

\hspace{\parindent}	Four years after Sommerfeld introduced his method, Hobson used it \cite{Gfc} to target the probleom of grounded conducting disk in 3D. Thirty-seven years later, Waldmann (a student of Sommerfeld) solved \cite{Zas} for Green functions using Sommerfeld's method by mapping half-plane to a disk of finite radius. After another thirty-four years, Davis and Reitz \cite{Spp} constructed Green functions for same problem using complex analysis.\\
As how we did in the previous chapter, first we build Laplace-Beltrami operator in $n$D to get Green functions after considering the required Dirichlet boundary conditions together with the corresponding discontinuities.\\
	Next we study the general characteristics of the harmonic functions for the $p$-norm wormholes. We focus on the Ellis wormhole case for $p=2$. Then we deform the wormhole such that the radii of those aligned hyperspheres become the Manhattan norm and see how Green functions behave after squashing the wormhole. Also we discuss $N=4$ case as $N=3$ is in the previously mentioned references. \\
	Finally we emphasize on the inversion transformation to the and see how to map Kelvin's method to Sommerfeld's method.\\

\section{Hyperspherical harmonic potentials and associated Green functions in $n$D curved space}
\hspace{\parindent}	For a charged point-particle located at the origin of $n$D Euclidean space, $\mathbb{E}_N$, the potential is governed by differential equation:
\begin{equation}\label{eq.3.1}
\nabla_{S^N}^2 \Phi_{\mathbb{E_N}}(\vec{r})= - \delta^N(\vec{r}) \, ,
\end{equation}
with Laplace-Beltrami operator in $S^N$ sphere \cite{Htf}: 
\begin{equation}\label{eq.3.2}
\nabla_{S^N}^2 = \frac{1}{r^{N-1}}\partial_r(r^{N-1}\partial_r)  - \frac{1}{r^2} L_{\mathbb{E}_N}^2 \, ,
\end{equation}
where $L^2$ (also $-r^2 \nabla_{S^{N-1}}^2$) is the angular derivatives:
\begin{equation}\label{eq.3.3}
L_{\mathbb{E}_N}^2= - r^2 \nabla_{S^{N-1}}^2 = \partial_{\theta}^2 + (N-2) \cot\theta \ \partial_{\theta} -r^2 \nabla_{S^{N-2}}^2 \, ,
\end{equation}
and $\theta$ represents the latitude from the $S^N$ hypersphere north pole. So for perceivable $S^2$ sphere:
\begin{equation}\label{eq.3.4}
\nabla_{S^{2}}^2 = \frac{1}{\sin^2\theta} \bigg[ \sin \theta \ \partial_{\theta}(\sin\theta \ \partial_{\theta})+\partial_{\phi}^2 \, \bigg] \, .
\end{equation}
Then for $N \geq 3$ the potential is:
\begin{equation}\label{eq.3.5}
\Phi_{\mathbb{E_N}}(\vec{r})= \frac{1}{(N-2) \ \Omega_N} \ \frac{1}{r^{N-2}}~ ,
\end{equation}
where the hyper solid angle $\Omega_N$ is given by:
\begin{equation}\label{eq.3.6}
\Omega_N = \frac{2\pi^{N/2}}{\Gamma(N/2)} ~\cdot
\end{equation}
The corresponding Green function is:
\begin{equation}\label{eq.3.7}
G_{\mathbb{E}_N} (\vec{r},\vec{r} \ ') = \frac{1}{(N-2) \ \Omega_N} \ \frac{1}{|\vec{r}-\vec{r} \ '|^{N-2}} ~ \cdot
\end{equation}
The second fraction in last equation can be expanded in terms of Gegenbauer polynomials $\mathcal{C}_l^{\frac{N-2}{2}}(\cos\theta)$ as:
\begin{equation}\label{eq.3.8}
\frac{1}{|\vec{r}-\vec{r} \ '|^{N-2}} = \sum \limits_{l=0}^{\infty} \frac{(r_<)^l}{(r_>)^{l+N-2}}\mathcal{C}_l^{\frac{N-2}{2}}(\hat{r}\cdot\hat{r} \ ') ~ ,
\end{equation}	

\begin{flushleft}
where $\displaystyle{\mathcal{C}_l^{\frac{N-2}{2}}(\cos\theta) = \sum \limits_{k=0}^{l/2}}\frac{(-1)^k(N+2l-2k-4)!!}{k!(2)^{l-k}(N-4)!!(l-2k)!!}(2\cos\theta)^{l-2k}$, with $r_>~, \ r_<~, \ \hat{r}~, \ \hat{r} \, '$ named as before and $\hat{r}\cdot\hat{r} \, '=\cos\theta$. In $N=3$ the Gegenbauer polynomials reduce to Legendre polynomials $\mathcal{P}_{lm}(\cos \theta)$.\\
\end{flushleft}

\noindent Now consider a generalized curved isotropic manifold with distance intervals:
\begin{equation}\label{eq.3.9}
(ds)^2 = (dw)^2+r^2(w)(d\hat{r})^2 \, ,
\end{equation}
where $\hat{r}$ represents the points on $S^{N-1}$, with $w \in (-\infty,+\infty)$, $\phi \in [0,2\pi]$ and\\ $\theta_i\in [0,\pi], \ \forall i=2, \cdots, N-2$.\\ Then $d\hat{r} = (\sin\theta_1)^{N-1} (\sin\theta_2)^{N-3}(\sin\theta_3)^{N-4}\cdots(\sin\theta_{N-2}) d\theta_1 d\theta_2 \cdots d\theta_{N-2} d \phi.$\\
The corresponding Laplace-Beltrami operator is:
\begin{equation}\label{eq.3.10}
\nabla_{S^N}^2 = \frac{1}{r(w)^{N-1}}\partial_w(r(w)^{N-1} \ \partial_w)  - \frac{1}{r(w)^2} L^2 \, .
\end{equation}
For 2D spherical harmonic eigenfunctions $Y_{lm}(\theta,\phi)=(-1)^m\sqrt{\frac{(2l+1)(l-m)!}{4\pi (l+m)!}} \mathcal{P}_{lm} (\cos\theta) \ e^{im\phi}$ of $L_{\mathbb{E}_2}^2$ on $S^2$, the $n$D hyperspherical harmonic eigenfunctions $Y_{lm_1m_2\cdots m_{N-2}}$ corresponding to $L_{\mathbb{M}_{N-2}}^2$ are \cite{Sph} :\\
\begin{equation}\label{eq.3.11}
Y_{lm_1m_2\cdots m_{N-2}} (\Omega) = \Bigg\{ \prod \limits_{i=1}^{N-3} A(\alpha_i,m_i) (\sin\theta_i)^{m_i} \ \mathcal{C}_{m_{i-1}-m_i}^{\alpha_i/2}(\cos\theta_i)\Bigg\} \times  \Bigg\{ Y_{m_{N-3}m_{N-2}} (\theta_{N-2},\phi) \Bigg\},
\end{equation}\\
where $\displaystyle{A(\alpha_i,m_i) = \Big[\frac{(2)^{\alpha_i -2} \ \Gamma^2(\alpha_i/2)  \ \Gamma(m_{i-1}-m_i+1) \ (\alpha_i+2(m_{i-1}-m_i))}{\pi \ \Gamma(\alpha_i+m_{i-1}-3m_i)} \Big]^{1/2}}$ and\\ 
$\alpha_i=2m_i+N-i+1$.\\
Other properties of spherical harmonics:
\begin{enumerate}
\item[i.] $L^2Y_{lm_1m_2\cdots m_{N-2}}=l(l+N-2)Y_{lm_1m_2\cdots m_{N-2}}$ \ .
\item[ii.] $\displaystyle{\int Y_{lm_1m_2\cdots m_{N-2}}(\hat{r}) \ Y_{lm_1m_2\cdots m_{N-2}}^{\star}(\hat{r} \, ') \ d\Omega = \delta_{ll'}\delta_{m_1m_1'} \cdots \delta_{m_{N-2}m_{N-2}'}}$ \ .

\item[iii.] $ \begin{aligned}[t] 
\sum \limits_{lm_1m_2\cdots m_{N-2}}Y_{lm_1m_2\cdots m_{N-2}}(\hat{r}) \ Y_{lm_1m_2\cdots m_{N-2}}^{\star}(\hat{r} \, ') &= \delta^{N-1}(\hat{r}-\hat{r} \, ')\\
& = \frac{2l+N-2}{(N-2)\Omega_N}\mathcal{C}_l^{\frac{N-2}{2}}(\hat{r}\cdot\hat{r} \, )~\cdot
\end{aligned}$
\end{enumerate}\label{it.3}

\noindent Using separation of variables we can write the harmonic functions on manifold with metric \eqref{eq.3.9} as:
\begin{equation}\label{eq.3.12}
h_{lm_1m_2\cdots m_{N-2}} =h_l(w) Y_{lm_1m_2\cdots m_{N-2}}(\hat{r}).
\end{equation}
Then the radial part of Poisson equation, which is \textit{homogeneous} differential equation, reads:
\begin{equation}\label{eq.3.13}
\begin{split}
\frac{1}{r(w)^{N-1}} & \partial_w \left(r(w)^{N-1} \ \partial_w h_l(w)\right)  = \frac{1}{r(w)^2} L^2 h_l(w) = \frac{l(l+N-2)}{r(w)^2}h_l(w) \, ,\\
\\
& \partial_w^2 h_l + \left[(N-1) \, \frac{r'(w)}{r(w)}\right] \partial_w h_l - \left[\frac{l(l+N-2)}{r(w)^2}\right]h_l(w)=0 \, .
\end{split}
\end{equation}
Then for such homogeneous differential equation with two linearly independent solutions $^{(1)}h_l$ and $^{(2)}h_l$, the Wronskian is:
\begin{equation}\label{eq.3.14}
\begin{split}
\mathbb{W}\left[^{(1)}h_l \, (w), \, ^{(2)}h_l \, (w)\right] = & \, ^{(1)}h_l \, (w) \frac{\overleftrightarrow{d}}{dw} \, {^{(2)}h_l} \, (w)\\
& = c_l \exp\left[-\int \limits_0^w (N-1) \frac{r'(\tilde{w})}{r(\tilde{w})} d\tilde{w} \right] = \frac{c_l}{r^{N-1}(w)}~,
\end{split}
\end{equation}
where $c_l= r^{N-1}(0) \times \mathbb{W}\left[^{(1)}h_l \, (0), \, ^{(2)}h_l \, (0)\right]$ is a constant.
For the examples of interest, $r(w) = r(-w)$. Therefore if $h_l(w)$ is a solution, then so is $h_l(-w)$. Also when $r(w) \substack{\\ \\ \approx \\ w \to \pm \infty} \, |w|$, then the solution of \eqref{eq.3.13} becomes:
\begin{equation}\label{eq.3.15}
^{(\pm)}{h_l(w)} \approx \Bigg\{ 
	\textstyle
    \begin{array}{c}
    \frac{1}{|w|^{l+N-2}} \, , \qquad w \to \pm \infty\\
    \\
    b_l \, |w|^l \, , \quad \qquad w \to \mp \infty
    \end{array} \, ,
\end{equation} 
where $^{(-)}{h_l(w)}= \, ^{(+)}{h_l(-w)}$. Notice in 3D space we have $\displaystyle{h_l\sim\left[\frac{1}{|w|^{l+1}}+|w|^l\right]}$ just like the well-known potential function of grounded conducting $S^2$ sphere.\\
So the Wronskian becomes:
\begin{equation}\label{eq.3.16}
\mathbb{W}\left[^{(-)}h_l \, (w), \, ^{(+)}h_l \, (w)\right] \substack{\\ \\ \approx \\ w \to + \infty} \, b_l \frac{2-N-2l}{w^{N-1}} ~ ,
\end{equation}
as expected from \eqref{eq.3.14} where $c_l=(2-N-2l)b_l~\cdot $\\

\noindent As expected, Green function must obey $G(\pm \infty ,\hat{r};w',\hat{r}')=0$ , together with discontinuity conditions and:
\begin{equation}\label{eq.3.17}
\nabla_{S^N}^2 G(w,\hat{r};w',\hat{r}') = - \frac{1}{r(w)^{N-1}} \ \delta(w-w \, ') \ \delta^{N-1}(\hat{r}-\hat{r}')~.
\end{equation}
From hyperspherical harmonics property \hyperref[it.3]{[iii]} we discussed earlier, and for $N>2$ dimensions, Green function (symmetric as usual under $w \leftrightarrow w'$) is:
\begin{equation}\label{eq.3.18}
\begin{split}
G(w,\hat{r};w',\hat{r}') & = \sum \limits_{lm_1m_2\cdots m_{N-2}} \frac{1}{c_l} \, ^{(+)}{h_l (w_>)} \,^{(-)}{h_l} \, (w_<) \, Y_{lm_1m_2\cdots m_{N-2}}(\hat{r}) Y_{lm_1m_2\cdots m_{N-2}}^{\star}(\hat{r} \, ')\\
& = \frac{1}{(N-2)\Omega_N} \sum \limits_{l=0}^{\infty} \frac{1}{b_l} \, ^{(+)}{h_l} (w_>) \, ^{(-)}{h_l} \, (w_<) \ \mathcal{C}_l^{\frac{N-2}{2}}(\cos\theta) \, ,
\end{split}
\end{equation}
where $c_l$ is given by the radial discontinuity:
\begin{equation}\label{eq.3.19}
\begin{split}
c_l = \lim \limits_{\varepsilon \to 0}\left[ \, ^{(+)}{h_l}(w') \frac{1}{dw} \left(r(w)^{N-1} \quad ^{(-)}{h_l}(w)\right) - \, ^{(-)}{h_l}(w') \frac{1}{dw} \left(r(w)^{N-1} \quad ^{(+)}{h_l}(w)\right) \right]_{w'=w+\varepsilon}
\end{split}
\end{equation}
Then the more general Green function that describes electrostatic potential function everywhere on both upper and lower branches for curved manifold with grounded innermost hypersphere with radius $R$ is:
\begin{equation}\label{eq.3.20}
G_o(w,\hat{r};w',\hat{r}')=G(w,\hat{r};w',\hat{r}')-G(w,\hat{r};-w',\hat{r}') \, ,
\end{equation}
with same comments on this function as those on $G_o$ at the end of both section \eqref{sec.2.4} and section \eqref{sec.2.2}.\\

The solution \eqref{eq.3.13} is obtained after doing the usual change of variable as in \eqref{eq.2.7} that renders $\partial_w = \frac{1}{r} \, \partial_u $. It guarantees $ r(w) \substack{\\ \\ \approx \\ w \to \pm \infty} \, |w| $ when $ u(w) \substack{\\ \\ \approx \\ w \to \pm \infty} \, \pm \infty $. So \eqref{eq.3.13} becomes:
\begin{equation}\label{eq.3.21}
\frac{d^2}{du^2}h_l + (N-2) r'(w(u)) \frac{dh_l}{du} = l(l+N-2)h_l \, \cdot
\end{equation}
Instead of writing $r'(w(u))$ as $\displaystyle{\frac{1}{r} \frac{dr}{du}}$ we keep it as it is because eventually we describe $r(w)$ radial function according to \eqref{eq.2.20}. But before we do that, \eqref{eq.3.21} is solved for arbitrary $r(w(u))$ by considering another change of variable $t=\tanh u, \ \text{for} \ t \in [1,1]$. Then \eqref{eq.3.21} becomes in form of Sturm-Liouville problem with its self-adjoint operator:
\begin{equation}\label{eq.3.22}
\frac{1}{(1-t^2)}\partial_t \left[(1-t^2) \partial_t \right] h_l + \frac{(N-2) \ r'(w(\tanh^{-1} t))}{(1-t^2)}\partial_t h_l = \frac{l(l+N-2)}{(1-t^2)^2} h_l ~,
\end{equation}
with corresponding integrating factor:
\begin{equation}\label{eq.3.23}
\mu (t) = \exp \left(\int \limits_0^t \frac{(N-2) \quad r'(w(\tanh^{-1} \tau))}{(1-\tau^2)} \ d\tau\right).
\end{equation}
Then \eqref{eq.3.22} becomes:
\begin{equation}\label{eq.3.24}
\frac{d}{dt} \left( \mu (t) \ (1-t^2) \frac{d}{dt} h_l \right) = \frac{l(l+N-2)}{(1-t^2)} \mu (t) \ h_l ~ ,
\end{equation}
which is determined based on the definition of $r(w(\tanh^{-1} t))$ in terms of variable $t$.

\section{$p$-norm and Ellis wormholes in $n$D}
\hspace{\parindent} Again for a class of $p$-norm radial functions defined as in \eqref{eq.2.20}, the asymptotic behavior of harmonics derived from \eqref{eq.3.13} is:
\begin{equation}\label{eq.3.25}
\frac{1}{|w|^{N-1}}\frac{d}{dw} \left( |w|^{N-1} \frac{d}{dw} h_l \right) \ \substack{\\ \\ \approx \\ |w| \gg R} \ \frac{l(l+N-2)}{|w|^2} h_l~\cdot
\end{equation}
In light of Gauss hypergeometric functions $\displaystyle{ \ _{2}{F}_1\left(\substack{a \ , \ b \\ \\ \\ c};z\right) = \sum \limits_{k=0}^{\infty} \frac{(a)_k \ (b)_k}{(c)_k} \frac{z^k}{k!}}$ where $\displaystyle{(a)_k=\frac{\Gamma(a+k)}{\Gamma(a)}}$ etc, it's more convenient \footnote{Like how we chose the solution of the \eqref{eq.2.7} to be logarithmic \eqref{eq.2.34}.} to express the previously used change of variable trick in terms of hypergeometric functions:
\begin{equation}\label{eq.3.26}
u(w)=\int\limits_0^w \frac{1}{\left(R^p+|(\tilde{w})^2|^{p/2}\right)} d\tilde{w}=\frac{w}{R} \ _{2}{F}_1\left(\substack{1/p \ , \ 1/p \\ \\ \\ 1+(1/p)} \ ; \ (\frac{w^2}{R^2})^{p/2}\right).
\end{equation}
Then for Ellis wormhole with $\displaystyle{u(w)=\ln\Big[ \frac{w}{R} + \sqrt{1+\frac{w^2}{R^2}}\Big]}$ and $r'(w(u))=\tanh u$, \eqref{eq.3.21} has two linearly independent solutions, where independency is guaranteed from the Wronskian we discussed before in \eqref{eq.3.14}.
\begin{equation}\label{eq.3.27}
\begin{split}
& \ ^{(1)}h_l(u) = \frac{(1+ \tanh u)^{(N+l-2)/2}}{(1-\tanh u)^{l/2}} \ _{2}{F}_1\left(\substack{2-(N/2) \ , \ (N/2)-1 \\ \\ \\ 2-(N/2)-l} \ ; \ \frac{1-\tanh u}{2}\right)  \, , \\
\\
& \ ^{(2)}h_l(u) = \frac{(1- \tanh u)^{(N+l-2)/2}}{(1+ \tanh u)^{l/2}} \ _{2}{F}_1\left(\substack{2-(N/2) \ , \ (N/2)-1 \\ \\ \\ (N/2)+l} \ ; \ \frac{1-\tanh u}{2}\right) \, .
\end{split} 
\end{equation}
It is crucial to take into consideration a linear combination of $ \ ^{(1)}h_l(u)$ and $ \ ^{(2)}h_l(u)$ as $ \ ^{(2)}h_l(u)$ behaves right for $N>2$ when $u \to +\infty$ , meanwhile $ \, ^{(1)}h_l(u) $ blows up  for odd $N$ upon $u \to -\infty$.\\

As we refered in the beginning of this chapter, harmonic potentials and their Green functions in 3D are discussed \cite{Gfc,Spp,Zas}. In 4D, harmonics are given by:
\begin{equation}\label{eq.3.28}
 \ ^{(\substack{1\\2})}h_l(u) = \frac{1}{\tanh u} e^{\pm (l+1)u} ~ , \qquad  \ ^{(\substack{1\\2})}h_l(\substack{-\infty \\ +\infty}) = 0 ~ ,
\end{equation}
such that \eqref{eq.3.14} becomes:
\begin{equation}\label{eq.3.29}
\mathbb{W}\left[^{(1)}h_l \, (u), \, ^{(2)}h_l \, (u)\right] = - \frac{2(l+1)}{\cosh^2 u} \, \cdot
\end{equation}
Upon changing the variable $t=\tanh{u}$, $(u=\mp\infty\to t=\mp 1)$, harmonic functions of \eqref{eq.3.27} become $^{(\substack{1\\2})}h_l(t=\mp 1)=0$, and \eqref{eq.3.29} becomes $\mathbb{W}\left[^{(1)}h_l \, (t), \, ^{(2)}h_l \, (t)\right] = - 2(l+1)$. So the linear independency of the two harmonics is still secured unless $l=-1$, which is non-physical from the definition of Gegenbauer polynomials \eqref{eq.3.8} where $l$ must start from 0. Therefore, the Wronskian never vanishes. This also guarantees that any other harmonic solution for same $l$ would strictly comprise of $^{(1)}h_l \, (t)$ and $^{(2)}h_l \, (t)$.\\

\noindent Back to Green function, in 4D case, we use $w_>$ and $w_<$ as before so \eqref{eq.3.18} renders:\\
\begin{equation}\label{eq.3.30}
\begin{split}
& G (w,\hat{r};w',\hat{r}')=\\
& = \frac{1}{4\pi^2 \sqrt{(R^2+w_>^2)(R^2+w_<^2)}} \sum \limits_{l=0}^{\infty} \Bigg[\frac{(w_> - \sqrt{R^2+w_>^2})(w_< + \sqrt{R^2+w_<^2})}{(w_> + \sqrt{R^2+w_>^2})(w_< - \sqrt{R^2+w_<^2})}\Bigg]^{\frac{l+1}{2}} \mathcal{C}_l^{(1)} (\hat{r}.\hat{r}')\\
& ~ \\
& =  \frac{1}{4\pi^2 \sqrt{(R^2+w_>^2)(R^2+w_<^2)}} \times \Bigg\{~\Big[\frac{(w_> - \sqrt{R^2+w_>^2})(w_< + \sqrt{R^2+w_<^2})}{(w_> + \sqrt{R^2+w_>^2})(w_< - \sqrt{R^2+w_<^2})}\Big]^{1/2}\\
& ~ \\
& \qquad\qquad\qquad\qquad\qquad\qquad + \Big[\frac{(w_> - \sqrt{R^2+w_>^2})(w_< + \sqrt{R^2+w_<^2})}{(w_> + \sqrt{R^2+w_>^2})(w_< - \sqrt{R^2+w_<^2})}\Big]^{-1/2} -2 (\hat{r}.\hat{r}')\Bigg\}^{-1}
\end{split}
\end{equation}
Asymptotically with $w,w'>0$, i.e., on the upper manifold branch, the last result is:

\begin{equation}\label{eq.3.31}
G(w,\hat{r};w',\hat{r}') ~ \substack{\\ \\ \approx \\ w,w' \gg R} ~  \frac{1}{4\pi^2 |\vec{r}-\vec{r} \, '|} +  \mathcal{O} \left( \frac{R}{r^3}, \frac{R}{r'^3} \right).
\end{equation}

\section{Manhattan norm in $n$D}
\hspace{\parindent} Studying ($p$=1)-norm of the radial function \eqref{eq.2.20} reveals the two flattened sheets of $\mathbb{E}_N$, where each sheet is indeed missing a $N$-ball of radius $R$, namely $\mathbb{B}_N (R)$. Then, as $p \to 1$, we get a single hypersphere $S^{N-1}$ \textquote{creasing} the two manifolds $\mathbb{E}_N - \mathbb{B}_N (R)$. Such hypersphere acts as \textquote{portal} or \textquote{doorway} between the identical copies of the Riemann double-space. As in \hyperref[ftnt6]{footnote (6)} of section (2.4) the opposite extreme, $p \to \infty$, for $n$D is endowed with an equatorial $S^{N-1}$ slice of the wormhole that turns out to be cylindrical tube of \cite{Viw} p. 488, Eqn(3).\\
With Manhattan norm \eqref{eq.2.36}, Green function \eqref{eq.3.18} becomes:
\begin{equation}\label{eq.3.32}
G (w,\hat{r};w',\hat{r}')=
\begin{dcases}
\frac{1}{(N-2)\Omega_N} \ \sum \limits_{l=0}^{\infty} \frac{(r(w_<))^l}{(r(w_>))^{l+N-2}} \ \mathcal{C}_l^{(\frac{N-2}{2})} (\hat{r}.\hat{r}') \ ,  \qquad \text{if both} \ \Big\{ \substack{w_> \\ \\ \\ w_<}>0\\[5pt]
\frac{1}{(N-2)\Omega_N} \ \sum \limits_{l=0}^{\infty} \frac{(R)^{2l+N-2}}{(r(w_<)(r(w_>))^{l+N-2}} \ \mathcal{C}_l^{(\frac{N-2}{2})} (\hat{r}.\hat{r}')~,~ \text{if both} \ \Big\{  \substack{w_> \\ \\ \\ w_<}\substack{>0 \\ \\ \\ <0}
\end{dcases}
\vspace*{0.5cm}
\end{equation}
It ensures the continuity of $G$ as $w_< \to 0$ where $r(w)$ is around $R$. Also $G\to 0$ as $w_{\gtrless} \to \pm \infty$.\\
Then for the case where both $w$ and $w'$ are on the same side from the branch cut of the double-space $\mathbb{E}_N - \mathbb{B}_N (R)$, i.e., the same wormhole hyperbranch, and in light of \eqref{eq.3.8} the first line of the last equation gives:
\begin{equation}\label{eq.3.33}
\begin{split}
G (w,\hat{r};w',\hat{r}') \vert_{\substack{w>0\\w'>0}} & = \frac{1}{(N-2)\Omega_N} \ \frac{1}{\left[r^2(w)+r^2(w')-2r(w)r(w') \, \hat{r}.\hat{r}'\right]^{\frac{N-2}{2}}}\\
& = \frac{1}{(N-2)\Omega_N} \ \frac{1}{|\vec{r}-\vec{r} \, '|^{N-2}} \, ,
\end{split}
\end{equation}
where if $w$ and $w'$ are on opposite branches, say $w>0$ and $w'<0$, then the second line of \eqref{eq.3.32} becomes:
\begin{equation}\label{eq.3.34}
\begin{split}
G (w,\hat{r};w',\hat{r}') \vert_{\substack{w>0\\w'<0}} & = \frac{1}{(N-2)\Omega_N} \ \frac{R^{N-2}}{\left[r^2(w)+r^2(w')+R^4 \, -2R^2 \, r(w) \, r(w') \, \hat{r}.\hat{r}'\right]^{\frac{N-2}{2}}}\\
& = \frac{1}{(N-2)\Omega_N} \ \frac{R^{N-2}}{\left[r^2(w) \, r^2(w') +R^4 -2R^2 \, \vec{r} . \vec{r} \, ' \right]^{\frac{N-2}{2}}} \, .
\end{split}
\end{equation}
Combining the last two equations gives the more general Green function:
\begin{equation}\label{eq.3.35}
%\begin{split}
G_o (w,\hat{r};w',\hat{r}') \vert_{\substack{w>0\\w'>0}} = \frac{1}{(N-2)\Omega_N} \ \left[ \frac{1}{|\vec{r}-\vec{r} \, '|^{N-2}} \, - \frac{1}{| \, \vec{r}-\vec{r} \, ' (\frac{R}{r \, '})^2 \, |^{N-2}} \left(\frac{R}{r(w')}\right)^{N-2} \right]~,
%\end{split}
\end{equation}
where such Green function shows antisymmetric behavior for $w'$ in each branch, i.e., $G_o (w,\hat{r};w',\hat{r}') \vert_{\substack{w>0\\w'>0}} = - G_o (w,\hat{r};w',\hat{r}') \vert_{\substack{w>0\\w'<0}}$.\\
And for $N=4$ the corresponding Green function is:
\begin{equation}\label{eq.3.36}
\begin{split}
G_o(\vec{r},\vec{r} \, ') = G_o(\vec{r} \, ' , \vec{r}) = \frac{1}{4\pi}\bigg[ \frac{1}{r^2(w)+r^2(w')-2\hat{r}.\hat{r}'}-\frac{R^2}{r^2(w) \, r^2(w') +R^4 -2R^2 \, \vec{r} . \vec{r} \, '} \bigg]~,
\end{split}
\end{equation}
where symmetry of Green function is maintained through $G_o(\vec{r},\vec{r} \, ') = G_o(\vec{r} \, ' , \vec{r})$, and contour plots of $G$ and $G_o$ are shown for the $N=4$ in Figures of \cite{Ghw}.

\section{Relating Kelvin and Sommerfeld images \\ by inversion in $n$D}
\hspace{\parindent} Inversion map we introduced in section \ref{sec.2.5} presents the position of the negative image, of the original charge at $\vec{r} \, '$, to be the \textit{reduced strength} at $ \vec{r} \, '' = \left(\frac{R}{r}\right)^{N-2} \vec{r} \, '$.\\
Also the Laplacian \eqref{eq.2.47}, which is not an eigenvector of such transformation \textit{expect} for 2D, maintains the asymptotic behavior of harmonic functions as in \eqref{eq.3.15}. In 2D harmonics become \textit{effectively} of order $r^l$ or $r^{-l}$, while for $(n\neq2)$D harmonics are of order $r^l$ or $r^{2-N-l}$.

\noindent According to \eqref{eq.2.48} and for $N\neq 2$ Green functions are \emph{not} invariant under inversion. However $G$ is still symmetric under $\mathfrak{r} \leftrightarrow \mathfrak{r}'$. Then \eqref{eq.2.49} becomes:
\begin{equation}\label{eq.3.37}
\begin{split}
\mathfrak{I} \big[\nabla_r^2 G(r,r')\big] & = \mathfrak{I} \big[-\frac{1}{\sqrt{g(r)}} \delta^{N}(\vec{r}-\vec{r} \, ')\big] \\
& \downarrow\\
\left[\frac{\mathfrak{r}^2}{R^2}\right]^2 \left[\nabla_{\mathfrak{r}}^2+\frac{2(2-N)}{\mathfrak{r}^2}D_{\mathfrak{r}}\right]  \left[\frac{(\vec{\mathfrak{r}})^{N-2} \, (\vec{\mathfrak{r}} \, ')^{N-2}}{R^{2N-4}}\, G(\vec{\mathfrak{r}},\vec{\mathfrak{r}} \ ') \right] & = - \frac{\mathfrak{r}^{N+1}}{R^{2N}} \, \delta^{N}(\mathfrak{r}-\mathfrak{r}') \, \delta^{N-1}(\hat{\mathfrak{r}}-\hat{\mathfrak{r}}').
\end{split}
\end{equation}
Last results map the lower branch of wormhole with Manhattan norm to the interior of innermost hypersphere while leaving the upper branched as it is.\\
In Ref.\cite{Ghw} we find a discussion with more details about how Green functions and corresponding Kelvin images of grounded conducting disk in 3D are constructed by Hobson \cite{Gfc}, by Waldmann \cite{Zas} (one of Sommerfeld students), and by Davis and Reitz \cite{Spp}. Despite neither of these projects bring up Riemannian geometry to the studies of the corresponding Green functions, it is easy to see that the inversion map \eqref{eq.3.37}, in $N=3$ case, relates Sommerfeld's method\textemdash we modified by incorporating Riemannian geometry into game\textemdash to complex variables analysis that is used in the previously mentioned studies.

\chapter{2D Grounded Elliptic Ring}

\hspace{\parindent} The problem of conducting ellipse (ellipsenfl\"{a}che) has been extinsively studied using complex methods since the $19^{\underline{th}}$ century \cite{Gfa}. However we still have freedom to choose another choice of real variables rather than the conventional complex plane to get the same results. We will show that for such coordinate choice, and for the sake of perceivable visual reasons, it is more advantageous to use Sommerfeld's method rather than Kelvin's one. Elliptic coordinates in real $xy$-plane with two foci on the $x$-axis at $\pm a$ are given by:
\begin{equation}\label{eq.4.1}
x=a \cosh u \cos v, \qquad y=a \sinh u \sin v, \qquad 0\leq u \leq \infty, \qquad 0\leq v \leq 2\pi \, .
\end{equation}
Then for a complex variable $z=x+iy=\Vert r\Vert e^{i\theta}, \, z \in \mathbb{C}$ with $r^2=x^2+y^2$ it can be reparameterized such that:
\begin{equation}\label{eq.4.2}
z=a \cosh (u+iv)~.
\end{equation}
Then Euler formula reads:
\begin{equation}\label{eq.4.3}
u+iv= \pm \cosh^{-1}(\frac{x+iy}{a}) + 2i\pi k, \qquad \text{for some} \, k \in \mathbb{Z} \, .
\end{equation}
By choosing the positive solution with $k=0$ we get:
\begin{equation}\label{eq.4.4}
u = \Re \mathfrak{e} \left(\cosh^{-1} \left(\frac{x+iy}{a}\right)\right), \qquad \qquad v = \Im \mathfrak{m} \left(\cosh^{-1} \left(\frac{x+iy}{a}\right)\right) \, .
\end{equation}
Now we construct the metric and Laplacian to get the corresponding Green function.

\section{Green functions for a 2D elliptic ring}

\hspace{\parindent} Building unique space intervals for 2D elliptic case will not change the fact that the Green function is for logarithmic potential. As Neumann emphasized on 2D electrostatic problems \cite{Heh}, we expect the ellipse to be exactly like that of the 2D ring. However we can double check this fact in the following:

\begin{equation}\label{eq.4.5}
\begin{split}
& dx  = (a \, \sinh u \, \cos v) du - (a \, \cosh u \sin v)dv \, , \\
& dy = (a \, \cosh u \, \sin v) du + (a \, \sinh u \cos v)dv \, , \\
& (dx)^2 + (dy)^2 = a^2 ({\sinh}^2 u + {\sin}^2 v) \bigg[ (du)^2+(dv)^2 \bigg] \, .
\end{split}
\end{equation}
Then the corresponding Laplacian is:
\begin{equation}\label{eq.4.6}
\nabla^2 = \frac{1}{a^2 ({\sinh}^2 u + {\sin}^2 v)} \left( \frac{\partial^2}{\partial u^2} + \frac{\partial^2}{\partial v^2} \right) \, ,
\end{equation} 
with potential function equation:
\begin{equation}\label{eq.4.7}
\nabla_{u,v}^2 \Phi (u,v) = -a^2 ({\sinh}^2 u + {\sin}^2 v) \ \frac{\varrho(u,v)}{\epsilon_0}~ \cdot
\end{equation}
Then, as expected, Green function obeys:
\begin{equation}\label{eq.4.8}
\left( \frac{\partial^2}{\partial u^2} + \frac{\partial^2}{\partial v^2} \right) G(u,v;u',v') = - \delta(u-u') \, \delta(v-v') \, ,
\end{equation}
with, as usual, \textquote{a} symmetric solution similar to \eqref{eq.2.18}:
\begin{equation}\label{eq.4.9}
G(u,v;u',v') = - \frac{1}{4\pi} |u-u'| - \frac{1}{4\pi} \ln \left(1+e^{-2|u-u'|}-2e^{-|u-u'|} \cos(v-v')\right) \, .
\end{equation}

\section{Kelvin's method}

\hspace{\parindent} For a grounded ellipse at the center and its points are at $u=U$ with $v\in [0,2\pi]$, the general Green function is:
\begin{equation}\label{eq.4.10}
G_o(u,v;u',v')=G(u,v;u',v')-G(u,v;(2U-u'),v')~,
\end{equation}
where the first $G$ is the fundamental Green function, while the second $G$ is the Green function copy produced by the image located at $u''=2U-u'$.\\
Before we discuss Green functions, it is worth accentuating we deal with mixed homogeneous Dirichlet and Neumann boundary conditions as: $$G(u,v;u',v') ~ \substack{\approx \\ u \to \infty} ~ \frac{1}{2\pi}(u'-U)~,$$ which means Green functions at infinity are not fixed. However $\frac{\partial}{\partial_u} G(u,v;u',v') ~ \substack{\approx \\ u \to \infty} ~ 0$, i.e., the second term of \eqref{eq.1.3} vanishes. So this problem \emph{in particular} is excepted to yield solutions under false impression that it is governed by Dirichlet boundary conditions alone, which is not true. The Neumann boundary conditions are still considered even if they do \emph{not} contribute to the solution.\\
For $u \leq U$ and $u' \leq U$, i.e., both charge and field locations are inside the grounded ellipse, the image is always outside the ellipse with $U\leq (2U-u'=u'') \leq 2U$ in a contrast with the location of image in $S^N$ spheres that is produced by \textit{inversion} mapping. Keeping in mind the ellipse is \textit{bona fide} ellipse not a circle, $a \neq 0, \, U \neq \infty$, the image is never at infinity, it is always located at $u' \in [U,2U]$.\\
For $u \geq U$ and $u' \geq U$, i.e., both charge and field locations are outside the grounded ellipse, with the charge outside the grounded ellipse that is not too far way from center($U \leq u'\leq 2U$), the image is inside the ellipse ($0 \leq (2U-u'=u'') \leq U$). The other case with $u' \geq 2U$ we have \textquote{negative} position of the image ($0 \geq 2U-u'$). This is not perceivable unless we consider that negative position to be in the \textit{second} sheet of Riemann surface, i.e., the image moves through the part of the semi-major axis (\textit{portal} or \textit{doorway}) connecting the elliptic foci to that opposite world. Also since ($0 \geq 2U-u' \geq 2U$) with respect to the original world, an observer in the second sheet would see the image as if it is the original charge with a location far way from the ellipse in that world ($u''|_{\text{upper sheet}} \mapsto u'|_{\text{lower sheet}}$ with $u'|_{\text{lower sheet}} \geq 2U|_{\text{lower sheet}}$).\\
So for  Kelvin images, to solve \eqref{eq.4.10} with real coordinates as \eqref{eq.4.1}, the \textit{real} interior of such 2D grounded elliptic conductor needs more interior extension for the case when a point-like charge is far from the conductor. Bird view figures visualizing different locations for sources and corresponding images are in \cite{Icr} Appendix. A.
\section{Sommerfeld's method}
\begin{figure}[H]
    \centering
        \includegraphics[width=\textwidth,height=\textheight,keepaspectratio]{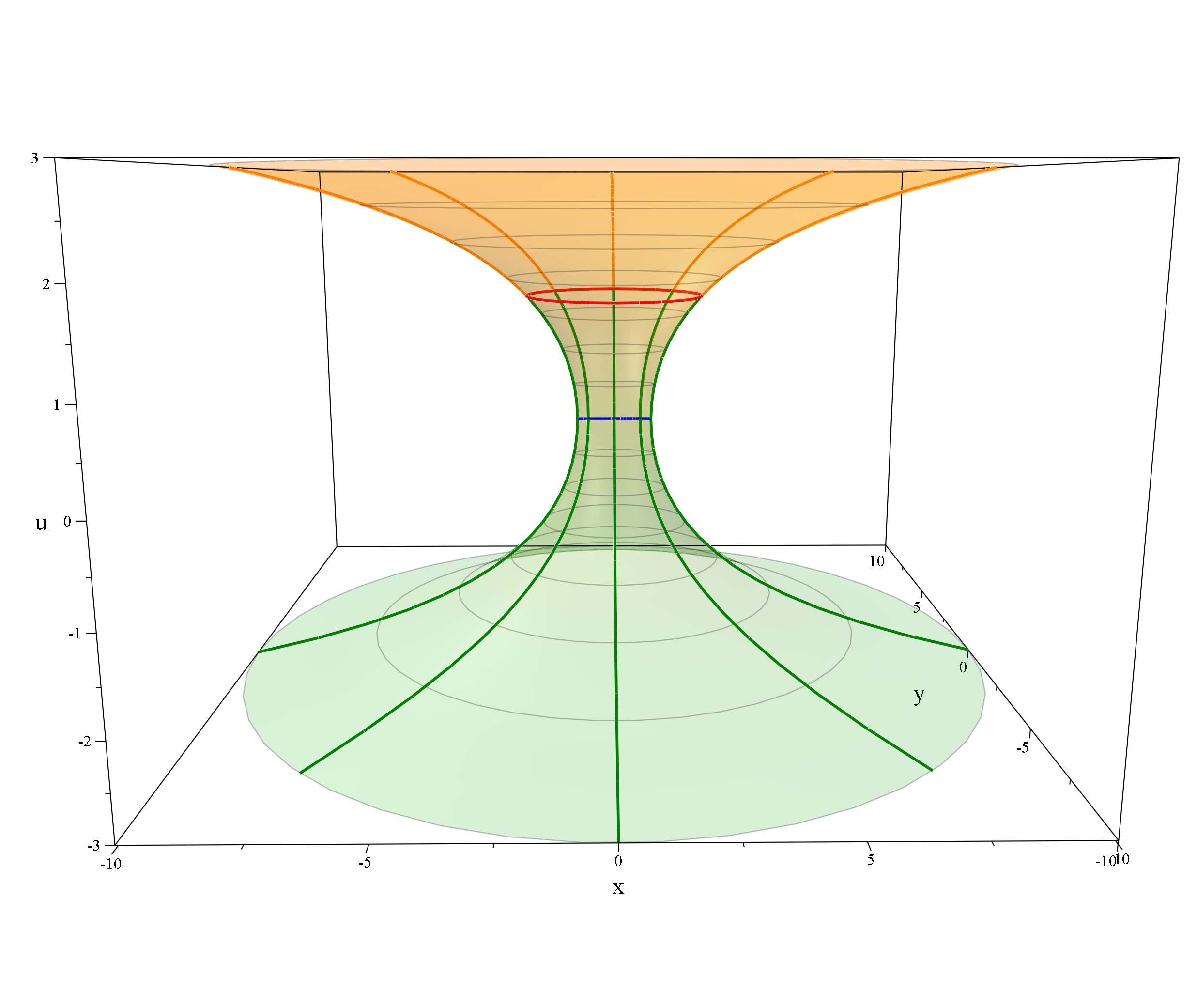}
        \caption*{Figure 1.a: Frontal view of extended real coordinates geometry to Kelvin's method for grounded ellipse (red) with $U=3/2$. The \textquote{doorway} (blue) is in the middle between  exterior charges (orange) and corresponding images (green) regions. Notice the phase change in the image trajectories $\Delta v=2 v'$ when they cross the ``doorway'' at $u=0$. Grey curves are parameterized ellipses with fixed $u$ and $0\leq v \leq 2\pi$.}
        \label{fig:KelFro}
\end{figure}

\begin{figure}[H]
        \includegraphics[width=\textwidth,height=\textheight,keepaspectratio]{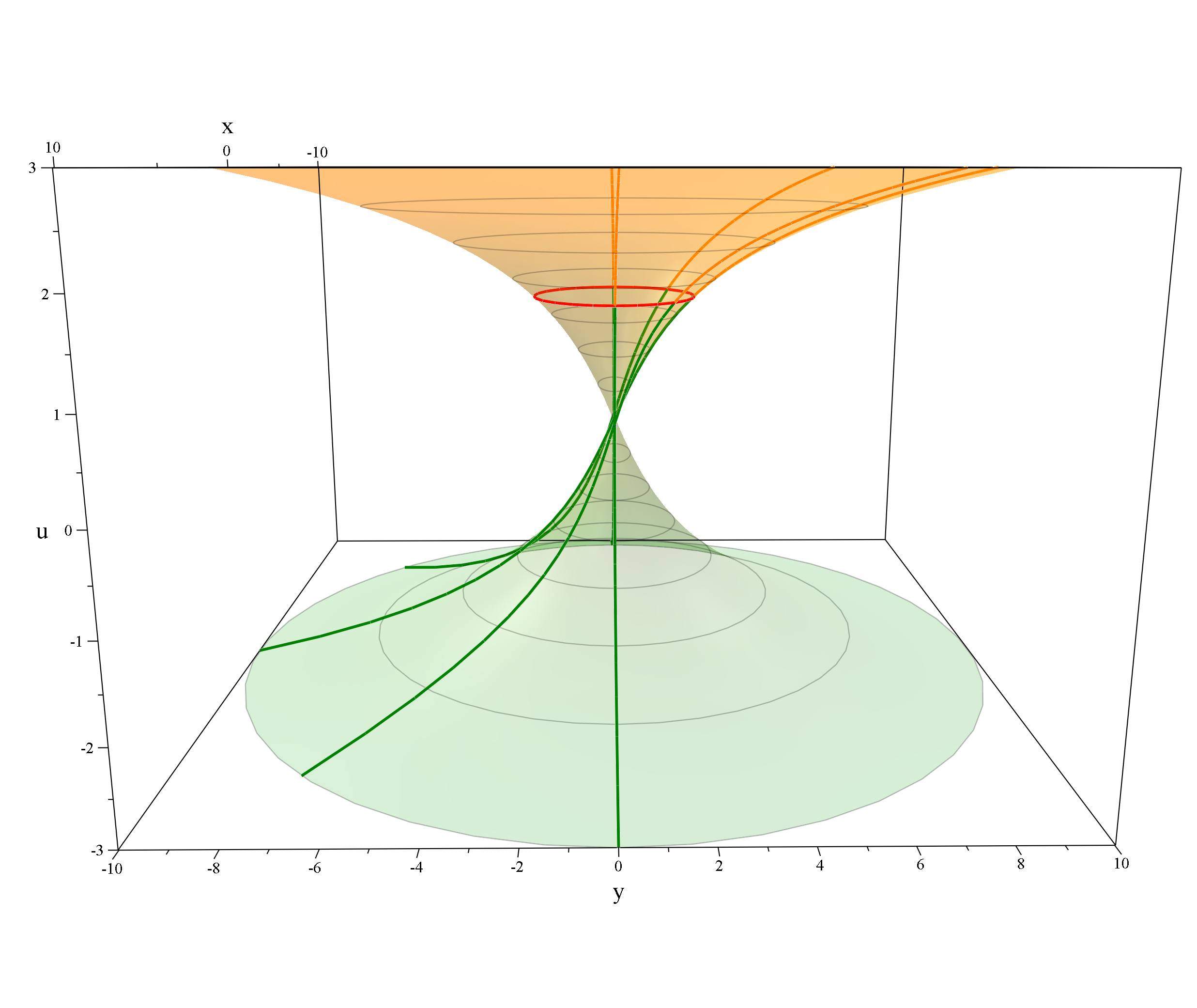}
        \caption*{Figure 1.b: Lateral view of of extended real coordinates geometry to Kelvin's method with for side view of grounded ellipse (red line) with $U=3/2$ where charges and images trajectories (orange) and (green) are shown respectively. Notice the phase change in the image trajectories $\Delta v=2 v'$ when they cross the ``doorway'' at $u=0$. Grey curves are parameterized ellipses with fixed $u$ and $0\leq v \leq 2\pi$.}
        \label{fig:KelLat}
\end{figure}

\begin{figure}[H]
    \centering
        \includegraphics[width=\textwidth,height=\textheight,keepaspectratio]{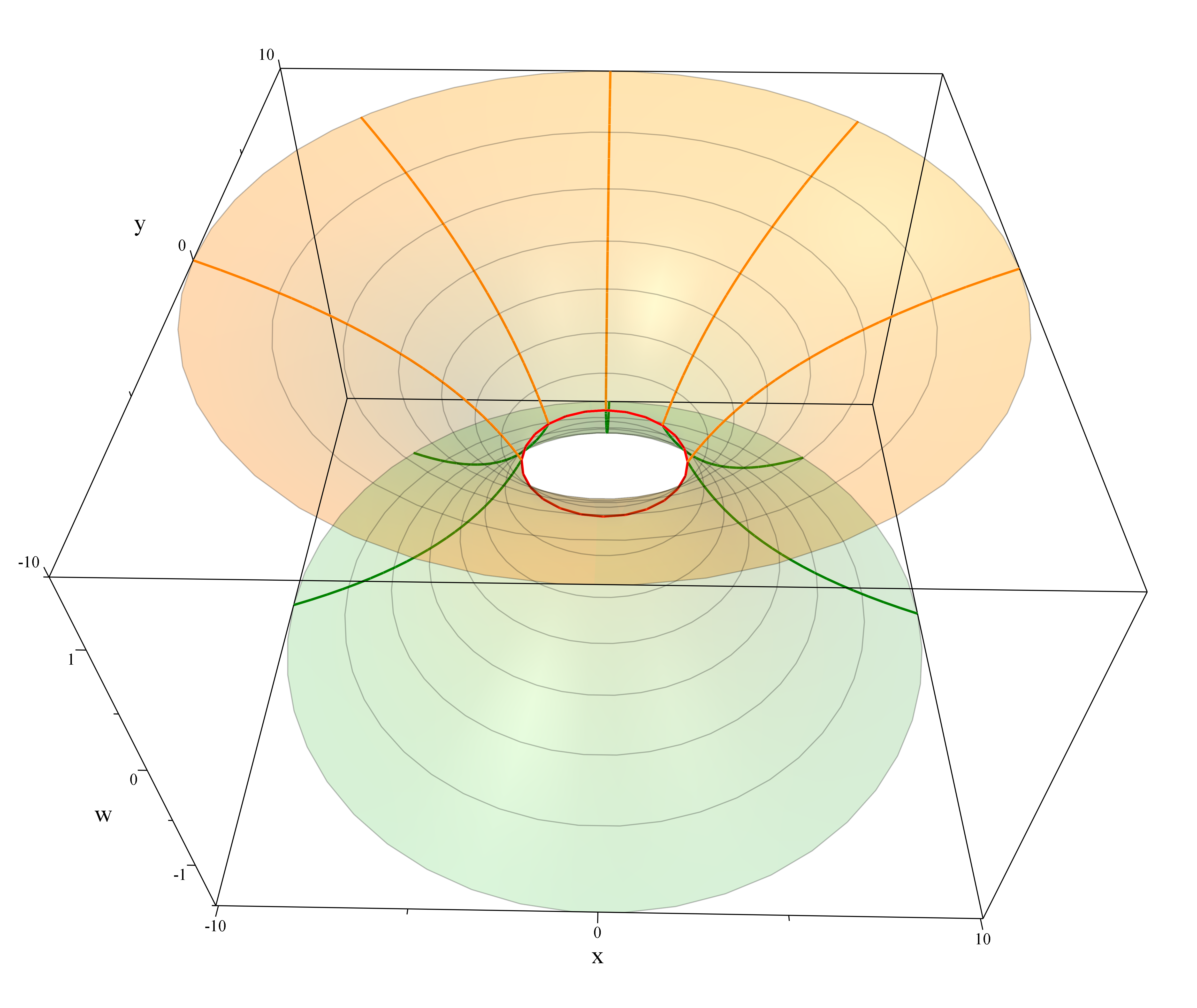}
        \caption*{Figure 2.a: Bird view of real coordinates geometry of Sommerfeld's method for grounded ellipse (red) at $U=0$ with trajectories of exterior sources (orange) and their corresponding images (green). In contrary to Kelvin's method, the ``doorway'' is the ring itself. Grey curves are parameterized ellipses with fixed $u$ and $0\leq v \leq 2\pi$.}
        \label{fig:SomTop}
\end{figure}

\begin{figure}[H]
    \centering
        \includegraphics[width=\textwidth,height=\textheight,keepaspectratio]{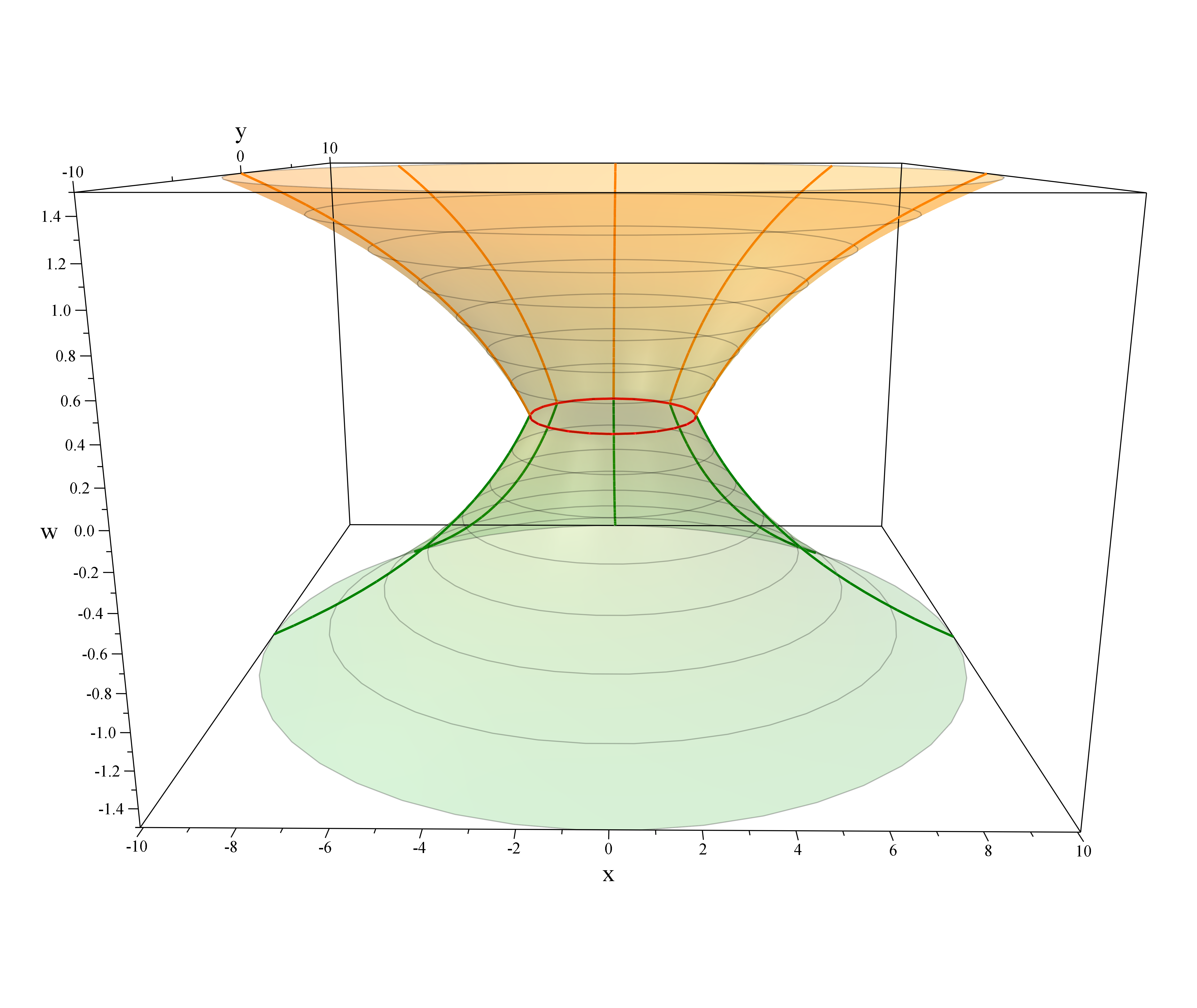}
        \caption*{Figure 2.b: Frontal view of real coordinates geometry of Sommerfeld's method for grounded ellipse (red) at $U=0$. Unlike Kelvin's method, Sommerfeld's method fixes trajectories of exterior sources (orange) and their corresponding images (green) at same angle $v$. Grey curves are parameterized ellipses with fixed $u$ and $0\leq v \leq 2\pi$.}
        \label{fig:SomFro}
\end{figure}

\hspace{\parindent}To avoid being repetitive, we jump immediately to the squashed wormhole case with Manhattan norm for the u parameter:
\begin{equation}\label{eq.4.11}
u=U+|w|~.
\end{equation}
For a reader uncomfortable with $du/dw$ discontinuity, we consider $u=(U^{2p}+w^{2p})^{1/2p}$ for $p\geq 1/2$ with $-\infty \leq w \leq \infty$ and repeat again what we discussed before.\\
Then for $0 \leq w, w' \leq + \infty$, \, \eqref{eq.4.9} becomes:
\begin{equation}\label{eq.4.12}
G(w,v;w',v') = - \frac{1}{4\pi} |w-w'| - \frac{1}{4\pi} \ln \left[1+e^{-2|w-w'|}-2e^{-|w-w'|} \cos(v-v')\right] \, .
\end{equation}
While for $0\leq w \leq \infty$ and $-\infty \leq {w'}_{lower} \leq 0$ Green function becomes:
\begin{equation}\label{eq.4.13}
\begin{split}
G(w,v;w_{lower}',v') & = G(w,v;-w_{up}',v')\\
& = - \frac{1}{4\pi} |w+{w'}_{up}| -\frac{1}{4\pi} \ln \left[1+e^{-2|w+{w'}_{up}|}-2e^{-|w+{w'}_{up}|} \cos(v-v')\right].
\end{split}
\end{equation}
By suppressing \textit{lower,up} and considering only $w,w' \in [0,+\infty)$, the more general Green function becomes:
 \begin{equation}\label{eq.4.14}
 \begin{split}
 G_o(w,v;w',v') & = G(w,v;w',v')-G(w,v;-w',v')\\
 & \begin{split} =
 & - \frac{1}{4\pi} \Big[|w-w'|+(w+w')\Big]\\
 & - \frac{1}{4\pi} \ln \left(1+e^{-2|w-w'|}-2e^{-|w-w'|} \cos(v-v')\right)\\
 & + \frac{1}{4\pi} \ln \left(1+e^{-2(w+w')}-2e^{-(w+w')} \cos(v-v')\right) \, .
 \end{split}
 \end{split}
 \end{equation}
Also notice $G(w,v;w',v')=G_o(w',v;w,v')$ with $G_o(-w,v;w',v')=-G_o(w,v;w',v')$ for positive $w$ and negative $w'$. For the sake of comparison, both Kelvin and Sommerfeld methods can be visualized in figures \hyperref[fig:KelFro]{(1.a)}, \hyperref[fig:KelLat]{(1.b)} and \hyperref[fig:SomTop]{(2.a)}, \hyperref[fig:SomFro]{(2.b)} respectively. Notice how parameterized curves behave when they pass by the ring from the perspective of each method.

\pagebreak

\section{Straight line limit}

\hspace{\parindent}As the focal distance $a \to \infty$, the elliptical coordinates $(u,v)$ in the region in between the two foci near $(u,v)\approx(0,\pi/2)$ become more rectangular Cartesian ones. This is achieved by applying these limits to \eqref{eq.4.1} so that $\lim \limits_{\substack{\\ a\to\infty \\ u\to 0}} (a \sinh u)=\frac{ay}{a}=y$ remains finite. 
Also when $x=0, \, v=\pi/2$ with $U=Y/a$, then $y(U/2,\pi/2) \to Y$. For such limits, the Green function \eqref{eq.4.9} becomes:
\begin{equation}\label{eq.4.15}
G(u=\frac{y}{a},v=\frac{\pi}{2};u'=\frac{y'}{a},v'=\frac{\pi}{2}) \, \substack{ \\ \\ \approx \\ a\to \infty} \, -\frac{1}{2\pi} \ln \left(\frac{|y-y'|}{a} + \mathcal{O}\left(\frac{1}{a}\right)\right) \, .
\end{equation}  
while \eqref{eq.4.10} becomes:
\begin{equation}\label{eq.4.16}
G_o(u=\frac{y}{a},v=\frac{\pi}{2};u'=\frac{y'}{a},v'=\frac{\pi}{2}) \, \substack{ \\ \\ \approx \\ a\to \infty} \, -\frac{1}{2\pi} \ln \left(\frac{|y-y'|}{|y-2Y+y'|} + \mathcal{O}\left(\frac{1}{a}\right)\right) \, .
\end{equation}
If $y,y' \geq Y$ then \eqref{eq.4.16} becomes:
\begin{equation}\label{eq.4.17}
G_o(u=\frac{y}{a},v=\frac{\pi}{2};u'=\frac{y'}{a},v'=\frac{\pi}{2}) \, \substack{ \\ \\ \approx \\ a\to \infty} \, -\frac{1}{2\pi} \ln \left(\frac{|y-y'|}{|y-2Y+y'|} + \mathcal{O}\left(\frac{1}{a}\right)\right) \, .
\end{equation}
This is exactly $G_o(y\geq Y,v=\frac{\pi}{2};y'\geq Y,v'=\frac{\pi}{2})$ for a grounded charge line ($y=Y$) parallel to the $x$-axis with Kelvin image at $(0,y'')=(0,2Y-y')$.\\
When $x\neq x'$ with the same $a \to \infty$ limit, the corresponding Green function of grounded half-plane is:
\begin{equation}\label{eq.4.18}
G_{\text{half \ plane}} (x,y;x',y') = - \frac{1}{2\pi} \ln \left(\frac{\sqrt{(x-x')^2(y-y')^2}}{\sqrt{(x-x')^2(y+y'-2Y)^2}}\right) \, ,
\end{equation}  
which is much more easy to derive using the complex variables rather that the real ones.
 
\section{Density of induced charge}

\hspace{\parindent} Since the density of the induced charge on the 2D ellipse is proportional to the normal component of the electric field near the ellipse $u \approx U$, it is known by considering the charge density induced on the 2D ellipse by an outsider point-like charge $u'\geq U$ to find the normal component of the electric field. This normal field is in Kelvin's method given by:
\begin{equation}\label{eq.4.19}
\mathbb{E}_{\mathbf{n}} (u) \Big|_{u=U} = -\frac{\partial G_o}{\partial u} \Big|_{u=U} \, ,
\end{equation}
or in Sommerfeld's method:
\begin{equation}\label{eq.4.20}
\mathbb{E}_{\mathbf{n}} (w) \Big|_{w=0} = -\frac{\partial G_o}{\partial w} \Big|_{w=0} \, .
\end{equation}
So in Sommerfeld's view the linear charge density for a point-like charge $(Q)$ at $w'>U$ is:
\begin{equation}\label{eq.4.21}
\lambda(v;w',v')= \mathbb{E}_{\mathbf{n}} (w) \Big|_{w=0} = -\frac{\partial G_o}{\partial w} \Big|_{w=0} = -\frac{1}{2\pi} \ \frac{e^{-2w'}-1}{e^{-2w'}-2e^{-w'} \cos(v-v')+1} \ Q~\cdot
\end{equation}
In the infinite limit, the last equation shows constant induced charge density on the ellipse:
\begin{equation}\label{eq.4.22}
\lambda(v;w',v') \ \substack{ \\ \\ \approx \\ w' \to \infty} \ -\frac{1}{2\pi} Q ~\cdot
\end{equation}
Notice total induced charge on the ellipse is:
\begin{equation}\label{eq.4.23}
\int_0^{2\pi} \lambda(v;w',v') dv = - Q  \, ,
\end{equation}
regardless wither $u' \approx U$ or $u' \to \infty$. \\

As our journey through wormholes is about to end, the last thing to be mentioned is that applying Sommerfeld's method to $n$D grounded ellipsoids comes with a major difficulty; there is \emph{no exact analytic} function that can describe Green functions for such case. The known Green functions so far are in form of infinite summations of \emph{ellipsoidal harmonics}. However, the principle stays the same, and therefore, upon finding an exact solution to Green functions in ellipsoidal cases, we will be able to construct and compare between both Kelvin and Sommerfeld images in different $p$-norms wormhole backgrounds. More on Green functions for $n$D ellipsoidal cases are discussed in \cite{Eht}.

\pagebreak
\chapter{Conclusion}
\hspace{\parindent} This thesis has focused on some mathematical aspects of Green functions using Sommerfeld's method for $p$-norm wormholes.\\

In chapter 1 we generally discussed Green functions and the necessary boundary conditions involved in solving for them. We heuristically presented the concept of Riemann surfaces and how to use it in combination with Sommerfeld's method to solve for Green functions of second order linear differential equations.\\

In chapter 2 we discussed the logarithmic potential for Poisson's equation and its corresponding Green functions in 2D curved surfaces. Inspired by the example of Ellis wormhole, we restricted the background to be a $p$-norm wormhole in preparation to obtain a Green function for a grounded ring. Then we squashed the wormhole into Manhattan norm and found the corresponding Green function for the grounded ring, along with all of its various properties. Later we related the Kelvin and Sommerfeld images and Green functions using inversion mapping, and we found the two viewpoints give mathematically equivalent Green functions.\\

In chapter 3 we investigated the same problem for grounded conducting hyperspheres in $n$D, and we compared and contrasted Green the functions of grounded conducting hyperspheres to that of the 2D grounded conducting ring.\\

In chapter 4 we analyzed the same problem using real variables to compare and contrast Kelvin and Sommerfeld methods for the 2D grounded conducting ellipse. We illustrated different ways of extending the interior region of the ellipse and placing images such that Dirichlet conditions are fulfilled. We also studied the straight line limit and showed how it exactly reproduced the Green function of a grounded half-plane. Then, we calculated the density of induced charge and proved that although the induced charge density itself behaves differently depending on where the original charge is, the total induced charge is the same as calculated using the conventional Kelvin's method. Finally, we referred to difficulties and suggestions on expanding this study to include ellipsoidal cases.\\

I hope this thesis has shown in principle that Sommerfeld's method simplifies the mathematics required to construct Green functions, in contrast with Kelvin's method, both conceptually and practically. I also hope that I have presented the not widely embraced mixture of Riemann surfaces and Green functions, a.k.a Sommerfeld's method, as a precursor toy model for the studies of wormholes.
 
\pagebreak

\begingroup
\setstretch{0.8}

\endgroup
\end{document}